\documentclass[useAMS,usenatbib]{mnras}
\pdfoutput=1 

\usepackage{graphicx}
\usepackage{setspace}
\usepackage{natbib}
\usepackage{color}
\usepackage{amsmath,amssymb}
\usepackage{times}

\usepackage{hyperref}[]

\def\aa{\rm \mathring{A}}

\usepackage{scalerel}
\usepackage{tikz}
\usetikzlibrary{svg.path}

\definecolor{orcidlogocol}{HTML}{A6CE39}
\tikzset{
  orcidlogo/.pic={
    \fill[orcidlogocol] svg{M256,128c0,70.7-57.3,128-128,128C57.3,256,0,198.7,0,128C0,57.3,57.3,0,128,0C198.7,0,256,57.3,256,128z};
    \fill[white] svg{M86.3,186.2H70.9V79.1h15.4v48.4V186.2z}
                 svg{M108.9,79.1h41.6c39.6,0,57,28.3,57,53.6c0,27.5-21.5,53.6-56.8,53.6h-41.8V79.1z M124.3,172.4h24.5c34.9,0,42.9-26.5,42.9-39.7c0-21.5-13.7-39.7-43.7-39.7h-23.7V172.4z}
                 svg{M88.7,56.8c0,5.5-4.5,10.1-10.1,10.1c-5.6,0-10.1-4.6-10.1-10.1c0-5.6,4.5-10.1,10.1-10.1C84.2,46.7,88.7,51.3,88.7,56.8z};
  }
}

\newcommand\orcidicon[1]{\href{https://orcid.org/#1}{\mbox{\scalerel*{
\begin{tikzpicture}[yscale=-1,transform shape]
\pic{orcidlogo};
\end{tikzpicture}
}{|}}}}

\title[Stochastic star formation and $z>10$ UV LF]{Stochastic star formation and the abundance of $z>10$ UV-bright galaxies}

\author[Kravtsov \& Belokurov]{Andrey Kravtsov$^{1,2,3}$\thanks{E-mail:kravtsov@uchicago.edu}\orcidicon{0000-0003-4307-634X} and  Vasily  Belokurov$^{4}$\orcidicon{0000-0002-0038-9584}\\
  $^1$Department of Astronomy and Astrophysics, The University of Chicago, Chicago, IL 60637 USA\\
  $^2$Kavli Institute for Cosmological Physics, The University of Chicago, Chicago, IL 60637 USA\\
  $^3$Enrico Fermi Institute, The University of Chicago, Chicago, IL 60637\\
  $^4$Institute of Astronomy, Madingley Rd, Cambridge, CB3 0HA, UK}

\begin{document}
\defcitealias{Aurora}{BK22}

\maketitle

\label{firstpage}

\begin{abstract}
We use a well-motivated galaxy formation framework to predict stellar masses, star formation rates (SFR), and ultraviolet (UV) luminosities of galaxy populations at redshifts $z\in 5-16$, taking into account stochasticity of SFR in a controlled manner. We demonstrate that the model can match observational estimates of UV luminosity functions (LFs) at $5<z<10$ with a modest level of SFR stochasticity, resulting in the scatter of absolute UV luminosity at a given halo mass of $\sigma_{M_{\rm UV}}\approx 0.75$. To match the observed UV LFs at $z\approx 11-13$ and $z\approx 16$ the SFR stochasticity should increase so that $\sigma_{M_{\rm UV}}\approx 1-1.3$ and $\approx 2$, respectively. Model galaxies at $z\approx 11-13$ have stellar masses and SFRs in good agreement with existing measurements. The median fraction of the baryon budget that was converted into stars, $f_\star$, is only $f_\star\approx 0.005-0.05$, but a small fraction of galaxies at $z=16$ have $f_\star>1$ indicating that SFR stochasticity cannot be higher. We discuss several testable consequences of the increased SFR stochasticity at $z>10$. The increase of SFR stochasticity with increasing $z$, for example, prevents steepening of UV LF and even results in some flattening of UV LF at $z\gtrsim 13$. The median stellar ages of model galaxies at $z\approx 11-16$ are predicted to decrease from $\approx 20-30$ Myr for $M_{\rm UV}\gtrsim -21$ galaxies to $\approx 5-10$ Myr for brighter ones. Likewise, the scatter in median stellar age is predicted to decrease with increasing luminosity.  The scatter in the ratio of star formation rates averaged over 10 and 100 Myr should increase with redshift. Fluctuations of ionizing flux should increase at $z>10$ resulting in the increasing scatter in the line fluxes and their ratios for the lines sensitive to ionization parameter. 
\end{abstract}


\section{Introduction}
\label{sec:intro}

The era of the James Webb Space Telescope (JWST) is revolutionizing our understanding of the earliest stages of galaxy evolution due to the unique capabilities of its instruments to probe stellar populations at $z>10$ \citep[e.g., see][for a review]{Robertson.2022}.
Indeed, deep observations with JWST during the first two years of its operation detected significant numbers of $z>9$ galaxies bright in the rest-frame ultraviolet (UV) band  \citep[e.g.,][]{Naidu.etal.2022, Castellano.etal.2022,Finkelstein.etal.2023,Harikane.etal.2023,Harikane.etal.2024,Donnan.etal.2023,Robertson.etal.2023,McLeod.etal.2024,Casey.etal.2024,Hainline.etal.2024}, which opened a window into the earliest stages of galaxy evolution. Deep surveys are also finding increasing numbers of fainter galaxies \citep[e.g.,][]{Hainline.etal.2024,Austin.etal.2023,Simmonds.etal.2024,Atek.etal.2023,Atek.etal.2024} which will lead to reliable measurements of the UV luminosity functions form (UV LFs) at these high redshifts. 

Abundance of bright galaxies was estimated in several recent studies \citep{Finkelstein.etal.2023,Wilkins.etal.2023,Robertson.etal.2024,Yung.etal.2024}, although robust measurement of redshift is often difficult and thus debate about the abundance of the brightest galaxies at $z>7$ is ongoing \citep[e.g.,][]{Desprez.etal.2023}. Comparisons of these estimates with model predictions led to the conclusions that theoretical models significantly underestimate the abundance of UV bright galaxies at $z>10$. 

Theoretical analyses indicated that to match observed estimates of the high-$z$ UVLF requires high star formation efficiency and suppressed feedback in the gas-rich, starbursting environments of these galaxies \citep{Boylan-Kolchin.2023,Dekel.etal.2023,Qin.etal.2023} or a top-heavy IMF \citep{Inayoshi.etal.2022,Trinca.etal.2024,Yung.etal.2024,Wang.etal.2023model,Qin.etal.2023,Ventura.etal.2024}.  At the brightest galaxy tail,  
\citet{Ferrara.etal.2023} suggest that a rapid decrease of dust obscuration may play a role in reconciling models with observations. 
Furthermore, the star formation rate (SFR) and UV luminosties of  $z>9$ galaxies may be overestimated due to the contribution of active galactic nuclei \citep[e.g.,][]{DSilva.etal.2023}.

At the same time, \citet{Keller.etal.2023}, \citet{McCaffrey.etal.2023}, \citet{Sun.etal.2023} and \citet{Kannan.etal.2023} showed that results of high-resolution galaxy formation simulations are generally consistent with observed abundance galaxies in the JWST surveys at $z\sim 10-12$ for the range of luminosities and star formation rates (SFRs) where simulations and observations overlap. This seemingly contradictory conclusion provides a hint of why some of the simpler models may have failed to match the abundance of the high-$z$ galaxies. Star formation in high-resolution simulations of galaxies at these redshifts tends to be very bursty, while simple models only account for the average star formation of galaxies of a given mass. 

Indeed, several studies used simple analytic models for star formation in dark matter halos to point out that strong UV variability that can be induced by the variations in the SFR and/or initial mass function (IMF) of stars can help explain the abundance of the UV-bright galaxies by up-scattering low-mass galaxies that happen to be UV-bright at a given time \citep{Ren.etal.2019,Mirocha.Furlanetto.2023,Mason.etal.2023,Shen.etal.2023,Munoz.etal.2023}. At the same time, \citet{Sun.etal.2023} showed that the SFR stochasticity in the FIRE-2 galaxy formation simulations is sufficient to produce UV LF consistent with measurements of the UV LF at $z\approx 10-12$. 

On the observational side, recent JWST observations of spectroscopically confirmed galaxies also indicate that star formation in galaxies of $\lesssim 10^9\, M_\odot$ at $z>5$ is very bursty \citep[e.g.,][]{Looser.etal.2023, Atek.etal.2023}, which is consistent with the inferences from the broad-band JWST NIRCam photometry modelling \citep[e.g.,][]{Dressler.etal.2024,Endsley.etal.2023,Simmonds.etal.2024,Asada.etal.2024,Ciesla.etal.2023,Cole.etal.2023}. These observational studies show that burstiness of star formation increases with increasing redshift and with decreasing galaxy luminosity. Thus, both theoretical models and observations that SFR is quite stochastic during the early stages of galaxy evolution. 

At the same time, it is not yet clear whether model predictions agree with observations in detail or what the physical drivers of the SFR stochasticity are. In galaxy formation simulations compelling mechanisms have been identified to explain the burstiness of SFR in dwarf galaxies. For example, \citet{Su.etal.2018} showed that feedback due to individual supernovae and hypernovae leads to significant bursts of SFR in the FIRE-2 simulations. \citet{Sugimura.etal.2024}, on the other hand, show that suppression of cooling due to H$_2$ dissociation by the FUV radiation delays star formation until gas accumulates and becomes gravitationally unstable, which leads to a strong burst. Subsequent removal of gas by the feedback-driven outflow leads to a lull in star formation until the cycle repeats. In both studies, however, the effects are specific to dwarf galaxies and were shown to be significant in halos of mass $M_{200}\lesssim 10^{9}\, M_\odot$. 

In higher mass halos bursty SFR can be driven by the imbalance between gas accretion time scale and time scales for star formation and feedback that prevents equilibrium star formation regime in the interstellar medium. \citet{Faucher_Giguere.2018} argued that all high-redshift galaxies should be in such a regime due to high accretion and merger rates expected during early stages of galaxy evolution. 

In this paper, we use a regulator-type galaxy formation model \citep{Kravtsov.Manwadkar.2022,Manwadkar.etal.2022} to explore the role of bursty star formation in shaping galaxy UV LFs at $z\sim 5-16$. The goal is to test and elucidate the conclusions of the previous studies discussed above. We show that our results are consistent with the results of \citet{Sun.etal.2023} who showed that the level of SFR stochasticity in FIRE-2 simulations is sufficient to match observed UV LFs at $z\approx 8-12$. Here we extend the analyses to a wider range of redshifts of $z\in 5-16$. We show that the amounts of stochasticity required to match observed UV LF at $z>10$ are smaller than deduced in some of the previous studies. At the same time, the stochasticity is required to increase with increasing redshifts from $\sigma_{M_{\rm UV}}\approx 0.75$ at $z<9$ to $\sigma_{M_{\rm UV}}\approx 1.-1.3$ at $z\approx 10-13$ and to $\sigma_{M_{\rm UV}}\approx 2$ at $z=16$. We further show that SFR stochasticity plays the key role in shaping galaxy UV LF. 

The paper is organized as follows.  In Section \ref{sec:methods} we describe the methodology used to construct a representative sample of halos at $z\in 5-16$ and galaxy formation model used to populate the halos with galaxies. We also present an approach we used to add stochasticity to the baseline model SFR in a controlled manner. We present our results in Section \ref{sec:results} and discuss them in Section \ref{sec:discussion}. Our results and conclusions are summarized in Section \ref{sec:conclusions}. In the appendix \ref{app:lf_taudep} we present a demonstration of how UV LF changes for different assumptions of star formation depletion time, while in the appendix \ref{app:mhacc} we present comparisons of the halo mass accretion rate approximation we use to model the evolution of halo mass in our framework to other calibrations. We show that our approximation is accurate across a wide redshift range of $0<z<20$ and that small differences in calibrations of the accretion rate at $z>10$ have a negligible effect on the predicted galaxy UV LF at $z\approx 12.$

In calculations throughout this paper we assume flat $\Lambda$+Cold Dark Matter ($\Lambda$CDM) cosmology with the mean density of matter in units of the critical density of $\Omega_{\rm m}=0.32$, mean density of baryons of $\Omega_{\rm b}=0.045$, Hubble constant of $H_0=67.11\,\rm km\,s^{-1}\,Mpc^{-1}$, the amplitude of fluctuations within tophat spheres of $R=8h^{-1}$ Mpc of $\sigma_8=0.82$, and the primordial slope of the power spectrum of $n_s=0.95$. Halo virial masses throughout this study are defined within the radius enclosing density contrast of 200 relative to the critical density at the corresponding redshift and are denoted as $M_{\rm 200c}$.

\section{Methods}
\label{sec:methods}

The galaxy formation framework we use in this study is applied to predict galaxy population properties for representative samples of model galaxies at all relevant luminosities and redshifts. This is done using samples of halos that follow the expected halo mass function at each considered redshift and halo mass evolution tracks constructed using an accurate approximation for the halo mass accretion rate, as described in Section \ref{sec:hevol}. 

The galaxy formation model used here was shown to reproduce observed properties of $\lesssim L_\star$ galaxies at $z=0$ down to the faintest ultra-faint dwarf galaxies \citep[][]{Kravtsov.Manwadkar.2022,Manwadkar.etal.2022,Kravtsov.Wu.2023} and is briefly outlined in Section \ref{sec:grumpy} below. The basic model is supplemented with a simple prescription for a controlled stochasticity in the star formation rate described in \citet[][see Section \ref{sec:sfr_stoch} below]{Pan.Kravtsov.2023} to explore its effects on the UV LF.

\subsection{Halo evolution model}
\label{sec:hevol}

To model the luminosity function of galaxies accurately at a given redshift $z$ we first construct large samples of model halos using the following procedure. We use an accurate cubic spline approximation of the cumulative halo mass function computed using \citet{Tinker.etal.2008} approximation of halo mass function and the inverse transform sampling method to generate a random sample of halo masses in a given volume. The halo sample thus accurately follows the \citet{Tinker.etal.2008} halo mass function by construction. Experiments showed that halo samples in the mass range $\log_{10}M_{\rm 200c}\in[9,13.5]$ and cubic volumes of size $L\in[500-2000]h^{-1}\,\rm Mpc$ provide an accurate representation of the abundance of halos hosting galaxies at $z\in [5-16]$. 

Large volumes are required to model the bright end of the galaxy luminosity function accurately. Given that the halo mass function is steep, halo samples over the entire mass range in such volumes are large. For example, the number of halos in the above mass range in a $L=500h^{-1}$ Mpc cube at $z=5$ is $\approx 1.74\times 10^8$ and this number is dominated by halos near the lower mass limit of the sample.  Modeling galaxy formation in all of the halos, however, is not necessary. Instead, we select a random fraction of halos for modeling which is given as a function of halo mass $M_{\rm 200c}$ using: 
\begin{equation}
    f(M_{\rm 200c}) = n \left(\frac{M_{\rm 200c}}{10^9\, M_\odot}\right)^\eta,
\end{equation}
where $n\approx 10^{-5}-5\times 10^{-6}$ and $\eta\approx 1.5-2.35$ provide sufficiently large samples of model galaxies to reliably measure luminosity functions at different $z\in [5,16]$. 
When the luminosity function is computed at a given $z$, each model galaxy is weighted by $f^{-1}(M_{\rm 200c})$. This procedure 
ensures good sampling of all relevant decades of the luminosity function, while keeping the number of model galaxies reasonably small. 

Once a halo sample is drawn at a given redshift, $z_{\rm f}$, we use each halo mass as a starting point and construct a halo mass evolution track. The track is computed by integrating the equation of halo mass evolution $\dot{M}_{\rm 200c}=\mu(M_{\rm 200c})$ back in time 
using an accurate approximation for the mean halo mass accretion rate $\dot{M}_{\rm 200c}$: 

\begin{equation}
\dot{M}_{\rm 200c} = 0.3606\, M_\odot\, {\rm Gyr}^{-1}\, M_{\rm 200c}(t)^{1.091}\, t^{-1.8},
\label{eq:dmdt}
\end{equation}
where halo mass $M_{\rm h}$ is in $M_\odot$ and time $t$ is in Gyrs. We present comparisons
of this simple equation to other approximations proposed in the literature, including the high-$z$ calibration of \citet{Yung.etal.2024b}, in the Appendix \ref{app:mhacc}. This approximation for mass accretion rate was derived using analyses of the mass evolution histories of halos formed in cosmological $\Lambda$CDM simulations. The comparisons show that the above approximation accurately describes the mean accretion rate of halos of a given mass and has a convenient simple form for integration in time. We also show in the Appendix \ref{app:mhacc} (see Figure \ref{fig:uvlf_z12_dmdt_comp} and related discussion) that small differences between different approximations have a negligible effect on the model UV luminosity function. 

Note that the equation \ref{eq:dmdt} approximates the {\it average} rate of accretion, while mass assembly histories of halos of a given mass have a scatter which corresponds to the scatter in the halo mass accretion rates. Although it would be straightforward to model this scatter, we choose not to do so in this study because the scatter in the star formation in our fiducial model is completely dominated by the scatter introduced by the stochastic SFR model described below. The scatter in the halo mass accretion rate produces a sub-dominant scatter of only $\lesssim 0.1$ dex in stellar mass at a given halo mass. 

In our analysis we thus consider two model variants: the base model that has no sources of SFR scatter and which describes only the average evolution of galaxy properties in halos of a given mass and the model in which different amounts of SFR stochasticity are introduced. 

\begin{figure*}
  \centering
  \includegraphics[width=\textwidth]{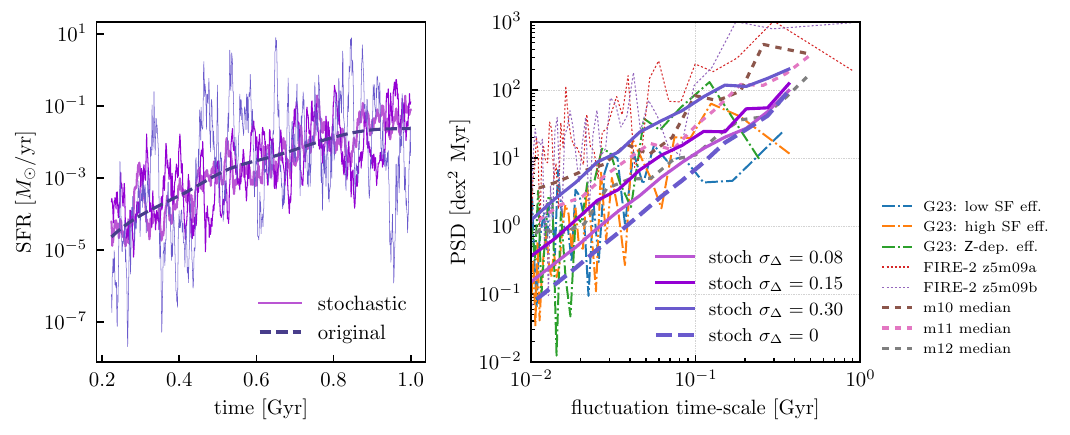}
  \caption[]{{\it Left panel:} Example star formation history of one of the model galaxies produced by the fiducial model (blue long dashed line) and star formation histories with different levels of stochasticity added ($\sigma_\Delta=0.08, 0.15, 0.30$ shown as violet, purple, and blue lines, respectively). {\it Right panel:} the median power spectrum densities of model galaxies in the fiducial model (thick blue long dashed line) and of the model star formation histories with different levels of stochasticity added (solid violet, purple and blue lines). The dotted, short-dash, and dot-dashed color lines show power spectra of $z>5$ galaxies in high-resolution zoom-in simulations by the FIRE-2 project \citep{Ma.etal.2018,Ma.etal.2019,Ma.etal.2020} and by \citet[][see text for details]{Garcia.etal.2023}.  Although the scatter is significant, there is a general trend of the PSD amplitude increasing with decreasing stellar mass of galaxies.}
   \label{fig:sfr_psd_sd008}
\end{figure*}
%

\subsection{The model of galaxy formation}
\label{sec:grumpy}

The halo mass evolution tracks computed as described above used as the basis for the \texttt{GRUMPY} galaxy formation model \citep{Kravtsov.Manwadkar.2022}  built upon a regulator-type galaxy formation framework \citep[e.g.,][]{Krumholz.Dekel.2012,Lilly.etal.2013,Feldmann.2013}. The model solves a system of differential equations that describe the evolution of gas mass, stellar mass, size, and stellar and gas-phase metallicities. It also includes galactic outflows, a model for the gaseous disk and its size, molecular hydrogen mass, star formation, and more. The model parameters used in this study are identical to those in the fiducial model of \citet{Manwadkar.etal.2022}, which was shown to reproduce properties of observed $L\lesssim L_\star$ galaxies at $z=0$ down to the ultra-faint galaxies \citep[cf. also][]{Kravtsov.Wu.2023}.

The \texttt{GRUMPY} model assumes that at all times the ISM follows the exponential radial gas profile, $\Sigma_g(R)$ with a half-mass radius proportional to the parent halo virial lradius. The mass of molecular hydrogen treated as a proxy for star-forming gas is estimated as 
\begin{equation}
    M_{\rm H_2} = 2\pi\int_0^{\rm{R_{HI}}}f_{\rm H_2}(\Sigma_g)\Sigma_g(R)RdR
\end{equation}
where $R_{\rm{HI}}$ is the radius corresponding to the assumed self-shielding threshold and $f_{\rm H_2}$ is the molecular fraction computed using current gas metallicity and UV flux proportional to the mean star formation rate using the model of  \citet{Gnedin.Draine.2014}. We note in passing that modeling of molecular gas at low metallicities is uncertain and star formation may not be traced by molecular gas \citep{Polzin.etal.2024}. We use the above prescription for molecular gas as a proxy for star-forming gas. As we noted, the model with this star formation prescription reproduces a broad range of properties of $L\lesssim L_\star$ galaxies and thus likely captures realistic scaling of the star-forming gas fraction with metallicity. 

The star formation rate is computed using $M_{\rm H_2}$ assuming a constant depletion time of molecular gas, $\tau_{\rm sf}$, and instantaneous recycling of the gas:
\begin{equation}
    \dot {\mathcal{M}}_\star = (1-\mathcal{R})\frac{M_{\rm H_2}}{\tau_{\rm sf}}
\end{equation}
where $\mathcal{R}=0.44$ is the gas fraction that is returned to the ISM via mass loss in winds and supernovae explosions assuming the \citet{Chabrier.2003} initial mass function of stars \citep[e.g.,][]{Leitner.Kravtsov.2011, Vincenzo.etal.2016}. We use the fiducial value of $\tau_{\rm sf}=2$ Gyr consistent with measurements in nearby galaxies \citep[e.g.,][]{Bigiel.etal.2008}, but we explore models in which $\tau_{\rm sf}$ decreases with increasing redshift, as indicated by observations of high $z$ galaxies. In addition, $\tau_{\rm sf}$ varies significantly from galaxy to galaxy. We do not model this scatter explicitly, because we will include different sources of scatter in SFR in a controlled manner using an explicit model of SFR stochasticity described below.  

The model includes outflows of the ISM gas that are assumed to be proportional to the mean SFR, $\dot{M}_{\rm out}=\eta_w M_{\rm H_2}/\tau_{\rm sf}$ with the mass-loading factor $\eta_w$ dependent on the current stellar mass of the galaxy in a way expected in the energy-driven wind models. With the outflow inclusion, the model successfully reproduces the general trends observed in the SFR--stellar mass relation, mass--metallicity relation, and several other properties of $<L_\star$ galaxies \citep{Manwadkar.etal.2022}. However, it does not include modeling of the processes that can result in significant SFR stochasticity, such as formation and destruction of individual star-forming regions \citep[e.g.,][]{Tacchella.etal.2020,Iyer.etal.2020,Sugimura.etal.2024}. As described above, we also do not introduce stochasticity due to scatter in halo assembly histories and depletion times. Instead, we introduce the collective SFR stochasticity due to these various processes in a controlled manner using the PSD formalism described below.

\subsection{The SFR stochasticity model}
\label{sec:sfr_stoch}

In addition to the evolution of galaxy properties computed in the base model, we model stochasticity of star formation relative to the mean SFR estimated by the base model by producing correlated Gaussian random number for each time step during galaxy evolution integration, as described in \citet{Pan.Kravtsov.2023}. The numbers are generated assuming a given power spectral density (PSD) of the time series in the Fourier domain. Namely, the mean SFR in the model at a time moment $t_n$ is perturbed as 
\begin{equation}
    {\rm SFR}_{\rm stochastic} = {\rm SFR}_{\rm mean}\times 10^\Delta,
\end{equation}
where $\Delta$ is a correlated random number drawn from the Gaussian pdf 
with zero mean and unit variance and multiplied by  $\sqrt{P(k)}=\sqrt{{\rm PSD}(f_k)/T}$, where wavenumber $k$ corresponding to frequency $f_k$ is defined as $k=f_k T$ and where $T$ is the duration of galaxy evolution track.

We follow  \citet{Caplar.Tacchella.2019} and use the PSD of the form
\begin{equation}
    {\rm PSD}(f) = \frac{\sigma^2_\Delta}{1+(\tau_{\rm break}f)^{\alpha}},
\label{eq:PSD}
\end{equation}
where $\sigma_\Delta$ characterizes the amplitude of the SFR variability over long time scales and  $\tau_{\rm break}$ characterizes the timescale over which the random numbers are effectively uncorrelated. Parameter $\alpha$ controls the slope of the PSD at high frequencies (short time scales). 

In our models, we fix the slope $\alpha$ and $\tau_{\rm break}$ to the values $\alpha=2$ and $\tau_{\rm break}=100$ Myr that are physically motivated by the time scales of gas evolution and star formation in giant molecular clouds in a typical ISM \citep[see][for a detailed discussion]{Tacchella.etal.2020}.
In the analyses presented below we only vary the amplitude of the PSD, $\sigma_\Delta$. Note that the SFR stochasticity is introduced after the model evolution is computed to predict the distribution of star formation rates and corresponding UV luminosities. It does not affect the outflows in the model that are assumed to be proportional to the mean star formation rate computed by the model. 

Physically, the stochasticity can be driven by fluctuations in gas mass accretion rate, feedback-driven outflows \citep{Tacchella.etal.2020}, and other modulation processes, such as far UV suppression of molecular gas formation in high-density regions \citep{Sugimura.etal.2024} and imbalance between gas accretion rate and the time scale of star formation and feedback in gas-rich galaxies \citep{Faucher_Giguere.2018}. 

The right panel of Figure \ref{fig:sfr_psd_sd008} shows examples of the PSDs of the SFRs modeled at $z>5$ using our galaxy formation model with stochasticity corresponding to different $\sigma_\Delta$ values with the thick solid (for $\sigma_\Delta>0$) and thick blue dashed (for $\sigma_\Delta=0$) line. The corresponding star formation histories are shown in the left panel of the figure. Note that the PSD for $\sigma_\Delta=0$ simply reflects the average shape of the star formation history ${\rm SFR}(t)$. The thin dotted and dot-dashed lines show PSDs of star formation histories of the individual galaxies in the FIRE-2 simulations of galaxies forming in halos $M_{200}=10^9\, M_\odot$ at $z=5$ \citep{Ma.etal.2018,Ma.etal.2019,Ma.etal.2020} and in galaxies at $z>7$ in the simulations of \citet[][]{Garcia.etal.2023}. The thin short-dashed lines show median PSDs of star formation histories of galaxies forming in halos
of mass $M=10^{10}\, M_\odot$,  $10^{11}\, M_\odot$,  $10^{12}\, M_\odot$ at $z=0$ \citep{Wetzel.etal.2023}. The left panel of the figure illustrates the variation of SFR for one of the model galaxies with these different levels of SFR stochasticity. 

Figure \ref{fig:sfr_psd_sd008} shows that the stochasticity of the SFR in these high-resolution zoom-in simulations can be well described by the model PSD described above with $\sigma_\Delta\approx 0.08-0.4$. In the analyses below will use the values in this range and will show that the stochasticity of this level is sufficient to match the observed UV LFs at $z\in [5,16]$.

\begin{figure}
  \centering
  \includegraphics[width=0.5\textwidth]{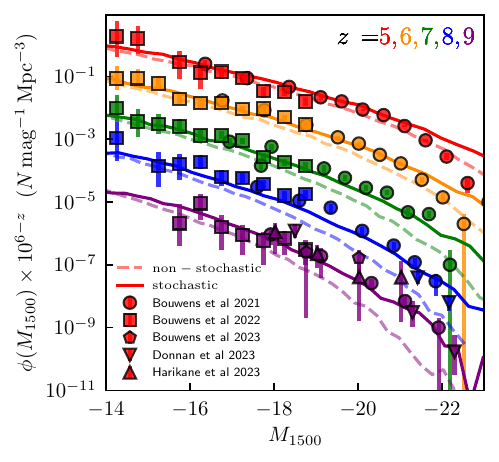}
  \caption[]{UV luminosity functions at redshifts $z=5, 6, 7, 8, 9$ (from the top down) measured from the HST and JWST observations (points) and predicted by the model (lines). LF at each redshift is offset by the factor of $10^{6-z}$ for clarity. The dashed lines show model UV LF with the fiducial parameters and depletion time $\tau_{\rm sf}=2$ Gyr when stochasticity is not included, while solid lines show results for the same model but with SFR stochasticity added with the PSD parameters $\sigma_\Delta=0.08$, $\alpha=2$, $\tau_{\rm break}=100$ Myr. }
   \label{fig:lf_z59}
\end{figure}

\subsection{Computing the UV luminosities of model galaxies}
\label{sec:muv}

To compute the monochromatic UV luminosity of model galaxies at $\lambda=1500\,\aa$, we tabulated the grid of luminosities, $L_{1500}$, for stellar populations of a given age and metallicity using the Flexible Stellar Population Synthesis model v3.0 \citep[FSPS,][]{Conroy.etal.2009,Conroy.etal.2010} and its Python bindings, \textsc{Python-FSPS}. The table was used to construct an accurate bivariate cubic spline approximation to compute $L_{1500}$ for stellar populations of a given age and metallicity produced by the model. 

Specifically, we use finely spaced time outputs of the models to compute the integral $L_{1500}$ due to all stars formed by the current time using the evolution of stellar mass and stellar metallicity computed by the model. We do not include any nebular emission in the calculation of $L_{1500}$. We also do not include dust effects, which are expected to affect the brightest galaxies in the UV at $z\lesssim 7$. At $z\gtrsim 7$ these effects are expected to be minor as indicated by the values of the UV spectral slope $\beta$ in observed galaxies and by theoretical models \citep[e.g.,][]{Dayal.etal.2022,Lewis.etal.2023} and are considerably smaller than the uncertainty of the UV luminosity function predictions due to uncertainty in the amount of SFR stochasticity.

\section{Results}
\label{sec:results}

\begin{table}
	\centering
	\caption{SFR stochasticity normalization and resulting scatter of $M_{\rm UV}$ in different model parameter choices at}
	\label{tab:sfr_stoch_params}
	\begin{tabular}{lcccr} 
		\hline\hline
		&$z=10.7$ & $z=12$ & $z=13$ & $z=16$\\
		\hline
            \multicolumn{5}{|c|}{$\tau_{\rm sf}=2$ Gyr, $\sigma_\Delta=0.08$, UV LF is not matched}\\
            \hline
		$\sigma_{M_{\rm UV}}$ & 0.69 & 0.83 & 0.84 & 0.85\\
            \hline
            \multicolumn{5}{|c|}{$\tau_{\rm sf}=2$ Gyr, variable $\sigma_\Delta$, UV LF is matched}\\
            \hline
		$\sigma_\Delta$ & 0.15 & 0.15 & 0.16 & 0.25\\
		$\sigma_{M_{\rm UV}}$& 1.1 & 1.4 & 1.5 & 2.2\\
            \hline
            \multicolumn{5}{|c|}{variable $\tau_{\rm sf}$, variable $\sigma_\Delta$, UV LF is matched}\\
            \hline
            $\tau_{\rm sf}$ (Gyr) & 1.0 & 0.5 & 0.5 & 0.25\\
		$\sigma_\Delta$ & 0.13 & 0.13 & 0.15 & 0.22\\
		$\sigma_{M_{\rm UV}}$& 0.97 & 1.2 & 1.4 & 1.9\\
		\hline
	\end{tabular}
\end{table}

\begin{figure*}
  \centering
  \includegraphics[width=\textwidth]{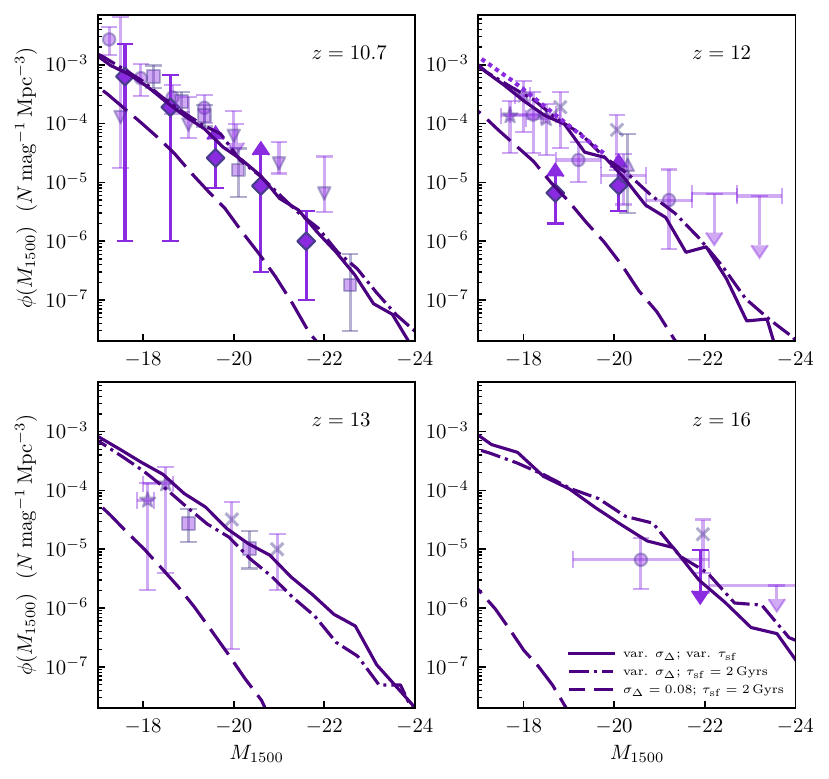}
  \caption[]{UV luminosity functions at redshifts $z=10.7, 12, 13, 16$ (from the top left panel clockwise) measured using JWST observations (points) and predicted by the model (lines). The dashed lines show model of UV LF with the fiducial parameters and decreasing depletion time $\tau_{\rm sf}=2$ Gyrs with stochasticity amplitude $\sigma_\Delta=0.08$ (PSD normalization), while dot-dashed lines show results for the same model but with $\sigma_\Delta$ increasing with increasing redshift from $\sigma_\Delta\approx 0.15$ at $z=10.7-13$ to $\sigma_\Delta\approx 0.25$ at $z=16$ (see Table \ref{tab:sfr_stoch_params}). The solid lines show a similar model but with star formation depletion time decreasing from $\tau_{\rm sf}=1$ Gyr at $z=10.7$ to $\tau_{\rm sf}=0.25$ Gyr at $z=16$; in this model, the amount of SFR stochasticity ($\sigma_\Delta$ values) required to match observed LFs are lower (see Table \ref{tab:sfr_stoch_params}). The dotted line in the $z=12$ panel (upper right corner) shows the UV LF estimate using the FIRE-2 simulations from \citealt{Sun.etal.2023}. The observational estimates using spectroscopic samples are shown by the brighter red, while the estimates from photometric samples are shown with fainter symbols. Specific measurements used at each redshift are: $z=10.7$  \citet[][diamonds]{Harikane.etal.2023,Harikane.etal.2024}, \citet[][squares]{Donnan.etal.2023}, \citet[][circles]{Leung.etal.2023}, \citet[][downward triangles]{Chemerynska.etal.2023}; $z=12$ \citet[][stars]{Robertson.etal.2024}, \citet[][crosses]{Bouwens.etal.2023}, \citet[][triangle]{Filkenstein.etal.2022}; $z=13$ \citet[][crosses]{Bouwens.etal.2023b}, \citet[][squares]{Donnan.etal.2023}, \citet[][stars]{Robertson.etal.2024}; $z=16$ \citet[][faint circle and upper limit]{Harikane.etal.2023}, \citet[][upper limit at $M_{\rm UV}=-21.9$]{Harikane.etal.2024}, \citet[][cross]{Bouwens.etal.2023b}. }
   \label{fig:lf_z1116}
\end{figure*}
%

\subsection{UV luminosity functions at $5<z<16$}
\label{sec:uvlf}

We first present the results of the galaxy formation model described above for the UV luminosity functions of galaxies at $z<10$ in Figure~\ref{fig:lf_z59}. The dashed lines show predictions of the model with star formation depletion time of $\tau_{\rm sf}=2$ Gyrs and without any sources of stochasticity of SFR, while the solid lines show the model with a moderate amount of stochasticity added to star formation rate with the power spectrum amplitude of $\sigma_\Delta=0.08$ (see eq. \ref{eq:PSD} in Section in \ref{sec:sfr_stoch}). This level of SFR stochasticity is comparable to the model shown by the magenta line (second lowest thick line) in the right panel of Figure \ref{fig:sfr_psd_sd008} and is on the low side of the stochasticity observed in high-resolution galaxy formation simulations shown in that figure. It corresponds to the scatter of $M_{\rm UV}$  at a given halo mass of $\sigma_{M_{\rm UV}}\approx 0.75$, which is comparable to $\sigma_{M_{\rm UV}}$ in high-resolution simulations at these redshifts \citep[][]{Pallottini.Ferrara.2023,Sun.etal.2023}. It is also consistent with the conclusions of \citet{Shen.etal.2023} about the amount of $M_{\rm UV}$ scatter required to reproduce UV LFs at $z\in[5,9]$.

Figure~\ref{fig:lf_z59} shows that the model with moderate SFR stochasticity provides a remarkably good match to observational measurements across the entire luminosity range probed by observations at these redshifts. Some minor discrepancies at luminosities $M_{1500}\lesssim -21$ and $z<7$ are likely due to dust attenuation not accounted in the model LF. 
As expected, the effect of the SFR stochasticity on the LF increases at brighter luminosities where LF is steeper. Likewise, the effect of stochasticity increases with increasing redshift for the same reason. 

Figure \ref{fig:lf_z1116} compares the same model with SFR stochasticity of  $\sigma_\Delta=0.08$ by the long-dashed lines and observational estimates of UV LF at $z>10$. At these redshifts, the model underpredicts the abundance of observed galaxies and the difference increases with increasing redshift. Such discrepancy of the model predictions and observations at these high redshifts is generic to most pre-JWST galaxy formation models \citep[e.g.,][]{Finkelstein.etal.2023,Harikane.etal.2023,Harikane.etal.2024,Wilkins.etal.2023,Yung.etal.2024,Chemerynska.etal.2023}. However, as we discussed above in Section \ref{sec:sfr_stoch}, the levels of stochasticity in very high-resolution simulations of galaxy evolution at $z>5$ is large and most previous models of UV LF do not take this into account. Indeed, Figure \ref{fig:lf_z1116} shows that if the SFR stochasticity amplitude is increased to $\sigma_\Delta=0.15-0.25$, comparable to those measured in the simulations of \citet{Ma.etal.2018,Ma.etal.2019,Ma.etal.2020} and \citet[]{Garcia.etal.2023}, the model (shown by the dot-dashed lines)
can match the observational estimates of UV LFs. 

\begin{figure}
  \centering
  \includegraphics[width=0.5\textwidth]{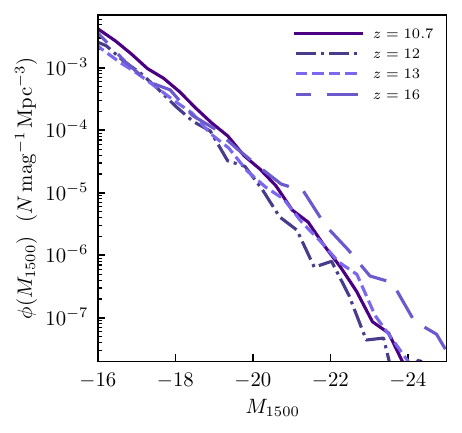}
  \caption[]{Model UV luminosity functions at redshift $z=10.7, 12, 13, 16$ (solid, dot-dashed, short-dashed, and long-dashed lines, respectively) shown side-by-side to illustrate the fact that increase of SFR stochasticity with increasing $z$ results in the LF that does not steepen at higher $z$ but becomes somewhat shallower at $z>12$.  }
   \label{fig:lf_flattening}
\end{figure}

Figure \ref{fig:lf_z1116} shows that the increase of SFR stochasticity with increasing $z$ required to match observations leads to the gradual flattening of the UV LF at higher $z$, which is a distinct {\it prediction} of this model. This can be seen in Figure \ref{fig:lf_flattening} which shows model UV luminosity functions at redshifts $z=10.7, 12, 13, 16$ side-by-side. It shows that the increase of SFR stochasticity with increasing $z$ results in the LF that does not steepen at higher $z$ but becomes somewhat shallower at $z>12$. 

The current observational measurements are consistent with such flattening \citep[see, e.g., Fig. 13 in][]{Harikane.etal.2023}, but measurements at $z\gtrsim 12$ are still too uncertain to probe the slope of the UV LF reliably. This prediction can be tested as the number of reliably detected galaxies at these redshifts increases. The model also predicts that the luminosity function extends to higher luminosities than the range probed by current observations. Future observations should thus find galaxies with higher UV luminosities than the maximum luminosities of the current samples. 

Note that the effect of SFR stochasticity on the model UV LF is twofold. The main effect is the upscattering of numerous lower-mass galaxies to higher luminosities due to SFR and UV luminosity bursts. Additionally, the lognormal scatter of SFR assumed in the model increases the mean SFR in galaxies of a given mass. The latter effect can be compensated by other parameters of the model affecting the average level of star formation. In particular, the depletion time of star-forming gas, $\tau_{\rm sf}$, directly affects the star formation efficiency, and thus SFR, for a given amount of gas. The solid lines in Figure \ref{fig:lf_z1116} show UV LFs of the model in which $\tau_{\rm sf}$ is allowed to be different at different $z$: specifically, $\tau_{\rm sf}=1$ Gyrs at $z=11$ and $z=12$, $\tau_{\rm sf}=0.5$ Gyrs at $z=11$ and $\tau_{\rm sf}=0.25$ Gyr at $z=16$. This model can produce UV LF similar to that of the model with $\tau_{\rm sf}=2$ Gyr, but with smaller SFR stochasticity. The specific normalization of the power spectrum of stochastic SFR required to match observed UV LF at each redshift is presented in Table \ref{tab:sfr_stoch_params}.

Although these depletion times values are smaller than the value of $\tau_{\rm sf}=2$ Gyr typical for $z=0$ galaxies \citep[e.g.,][]{Bigiel.etal.2008}, they are comparable to those observed in starbursting galaxies \citep[e.g.,][]{Kennicutt.Evans.2012,Huang.Kauffmann.2014,Hunt.etal.2015,Diaz.Garcia.Knappen.2020} and generally in galaxies at higher $z$ \citep[e.g.,][]{Tacconi.etal.2020}. In particular, direct estimates of molecular depletion time in galaxies at $z\gtrsim 6$  indicate values of measured $\approx 0.05-1$ Gyr \citep{Pavesi.etal.2019,Vallini.etal.2023}. 

Nevertheless, the effect of decreasing $\tau_{\rm sf}$ alone on the UV LF is insufficient. As we show in Appendix~\ref{app:lf_taudep}, a factor of ten decrease of depletion time does increase the amplitude of LF by an order of magnitude but this is still two orders of magnitude short of matching the observational estimates of LF amplitude at this redshift. Thus, in our model LF at $z\gtrsim 13$ cannot be matched by a realistic increase of star formation efficiency. This is because the increase of star formation rate is partly offset by the increase in the gas outflow rate, which is proportional to the SFR.

\subsection{UV luminosity--stellar mass relation of $z>10$ model galaxies}
\label{sec:sfrs}

\begin{figure*}
  \centering
  \includegraphics[width=\textwidth]{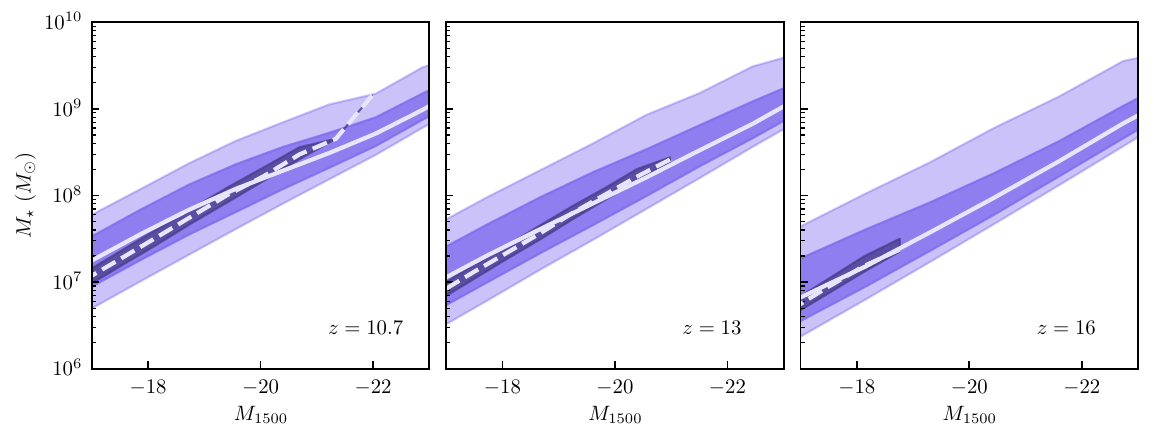}
  \caption[]{UV absolute magnitude--stellar mass correlation of model galaxies at $z=10.7, 13, 16$. The dashed lines show the relation in the model without SFR stochasticity. The solid lines show the median $M_\star$ in bins of $M_{1500}$ in the model with SFR stochasticity that matches observed UV LF at $z>10$ (shown by solid lines in Fig. \ref{fig:lf_z1116}).  The shaded regions show the $68.2$ and $95.4$ percentiles of the stellar mass distribution.}
   \label{fig:m1500_mstar}
\end{figure*}

Figure \ref{fig:m1500_mstar} shows the UV absolute magnitude--stellar mass correlation of model galaxies at $z=10.7, 13, 16$ in models with and without SFR stochasticity. The median correlation can be described approximately by a relation $\log_{10}M_\star = \log_{10}M_{-20\star} + s(M_{1500}+20)$, where $\log_{10}M_{-20\star}=8.2, 8.04, 7.8$ and $s=0.25, 0.33, 0.4$ at $z=10.7, 13, 16$, respectively. The scatter around the correlation in the model with SFR stochasticity is moderate because scatter in stellar mass and absolute magnitude are correlated. Nevertheless, the scatter is both non-negligible and asymmetric which can make a conversion of the UV luminosity to $M_\star$ subject to biases unless a correct form of the pdf $p(M_\star\vert M_{1500})$ is adopted and scatter is taken into account properly. 

\begin{figure*}
  \centering
  \includegraphics[width=0.49\textwidth]{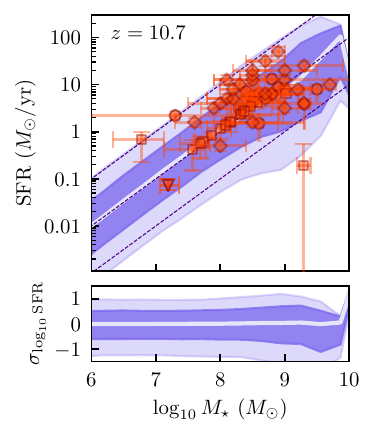}
  \includegraphics[width=0.49\textwidth]{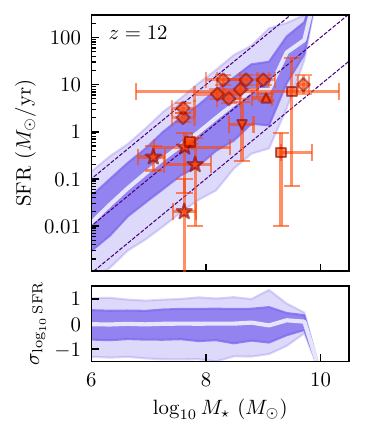}
  \includegraphics[width=0.49\textwidth]{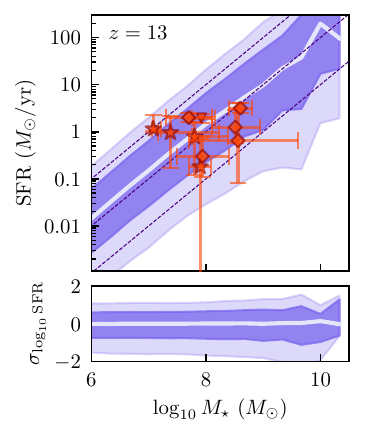}
  \includegraphics[width=0.49\textwidth]{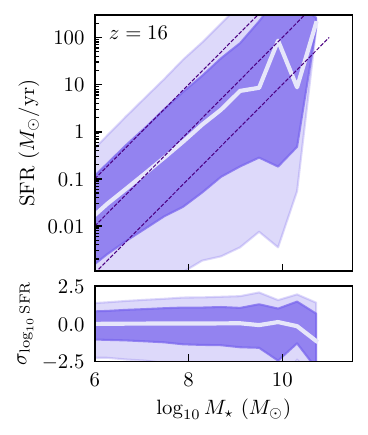}
  \caption[]{Star formation rate as a function of stellar mass in model and observed galaxies at $z=10.7, 12, 13, 16$. The solid lavender lines show the median relations for the fiducial model but with SFR stochasticity added with the PSD parameters  $\alpha=2$, $\tau_{\rm break}=100$ Myr and $\sigma_\Delta=0.13, 0.13, 0.15, 0.22$ at $z=10.7, 12, 13, 16$. The UV LF for this model is shown by the solid lines in Figure~\ref{fig:lf_z1116}. The $1\sigma$ and $2\sigma$ scatter in SFR at a given $M_\star$ is shown by the dark and light blue bands.
  The three dotted lines show linear relation ${\rm SFR}={\rm sSFR}\times M_\star$ for ${\rm sSFR}=10^{-7}, 10^{-8}, 10^{-9}\,\rm yr^{-1}$.
  The lower subpanels in each of the four redshift panels show the $1\sigma$ and $2\sigma$ scatter in $\log_{10}\rm SFR$ around the median. The red points in the first three panels show available observational measurements for galaxies at the corresponding redshifts (squares, \citealt{Atek.etal.2023}; triangle in the $z=10.7$ panel, \citealt{Roberts.Borsani.etal.2023}; circles, \citealt{Franco.etal.2023}; stars, \citealt{Robertson.etal.2024}; upward triangle in the $z=12$ panel, \citealt{Castellano.etal.2024}; downward triangles in the $z=12$ and $z=13$ panels, \citealt{Wang.etal.2023}); diamonds, \citealt{Morishita.etal.2024} and \citealt{Morishita.Stiavelli.2023}).}. 
   \label{fig:ms_sfr_z1116_dec_tauh2}
\end{figure*}

\subsection{Star formation rates of $z>10$ model galaxies}
\label{sec:sfrs}

Figure \ref{fig:ms_sfr_z1116_dec_tauh2} shows the evolution of the $M_\star$-SFR relation from $z=16$ to $z=11$ in the model with variable $\sigma_\Delta$ and $\tau_{\rm sf}$. At $z=11-13$ we also plot star formation rates and stellar masses inferred for individual observed galaxies at the corresponding redshifts. 
The figure shows that the model with stochastic SFR is broadly consistent with observations, which means that model galaxies of a given stellar mass have correct star formation rates. It also shows that matching UV LFs at $z>10$ requires that the scatter of SFR at higher $z$ increases dramatically with increasing redshift. At $z=10.7$ the scatter in $\log_{10}\rm SFR$ at a given $M_\star$ is $\approx 0.4-0.5$ dex, which is consistent with a recent observational estimate at that redshift by \citet{Cole.etal.2023}. It remains to be seen if the required scatter in the SFR is consistent with observations at higher $z$. As noted above for the UV LF, the model predicts the existence of galaxies with larger stellar masses and SFRs than in the population detected by the current surveys. These galaxies are rare but should be found  
as the volume probed by surveys increases. 

\begin{figure*}
  \centering
  \includegraphics[width=\textwidth]{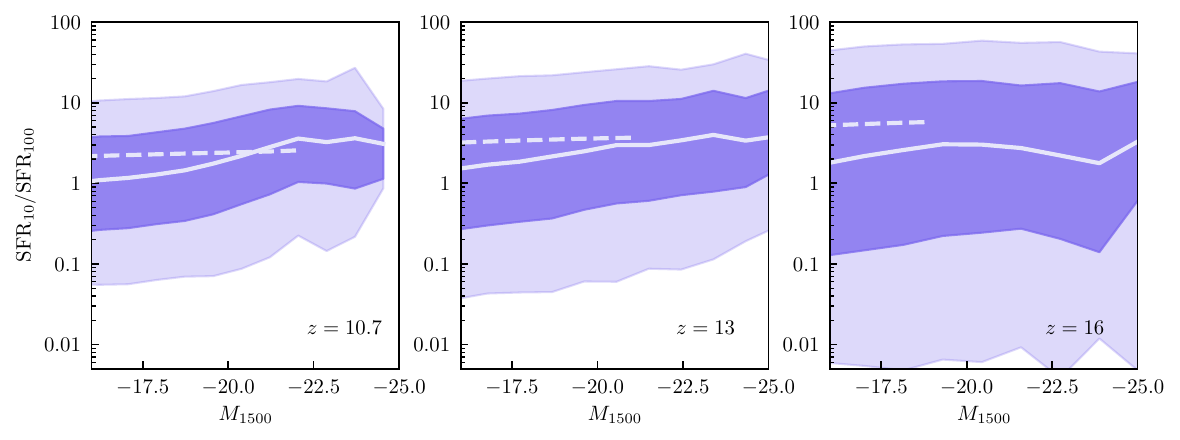}
  \caption[]{The ratio of star formation rates averaged over the past 10 Myr and 100 Myr as a function of the absolute magnitude at $1500\, {\rm \AA}$ for model galaxies at $z=10.7$, $z=13$, $z=16$. The dashed lines show the median $\rm SFR_{10}/SFR_{100}$ in the model without SFR stochasticity, while solid lines show the median for the fiducial model with SFR stochasticity at the level that matches observational estimates of UV LF at these redshifts. The shaded bands show the $68\%$ and $95\%$ of the distribution for the latter model. }. 
   \label{fig:sfr10_scatter}
\end{figure*}

\begin{figure*}
  \centering
  \includegraphics[width=\textwidth]{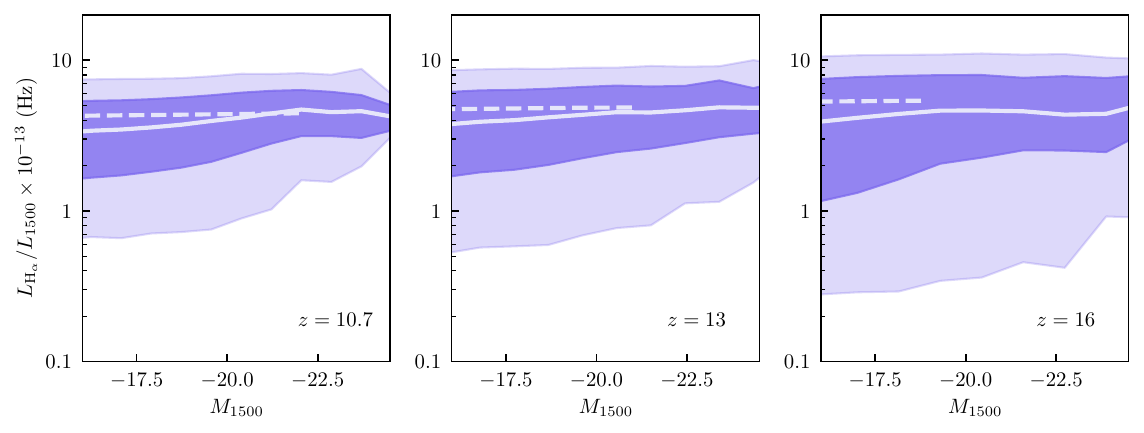}
  \caption[]{The ratio of the H$_\alpha$ luminosity to the luminosity per unit frequency at $\lambda=1500\,\rm \AA$ as a function of the absolute magnitude at $1500\, {\rm \AA}$ for model galaxies at $z=10.7$, $z=13$, $z=16$. H$_\alpha$ luminosity is proportional to the flux of hydrogen ionizing photons produced by the youngest stellar populations of model galaxies (age $\lesssim 3-5$ Myr), while UV luminosity reflects star formation history on the time scale of tens of Myr. The dashed lines show the median ratio in the model without SFR stochasticity, while solid lines show the median ratio for the fiducial model with SFR stochasticity at the level that matches observational estimates of UV LF at these redshifts. The shaded bands show the $68\%$ and $95\%$ of the distribution for the latter model.}
   \label{fig:lha_luv_scatter}
\end{figure*}

\begin{figure*}
  \centering
  \includegraphics[width=\textwidth]{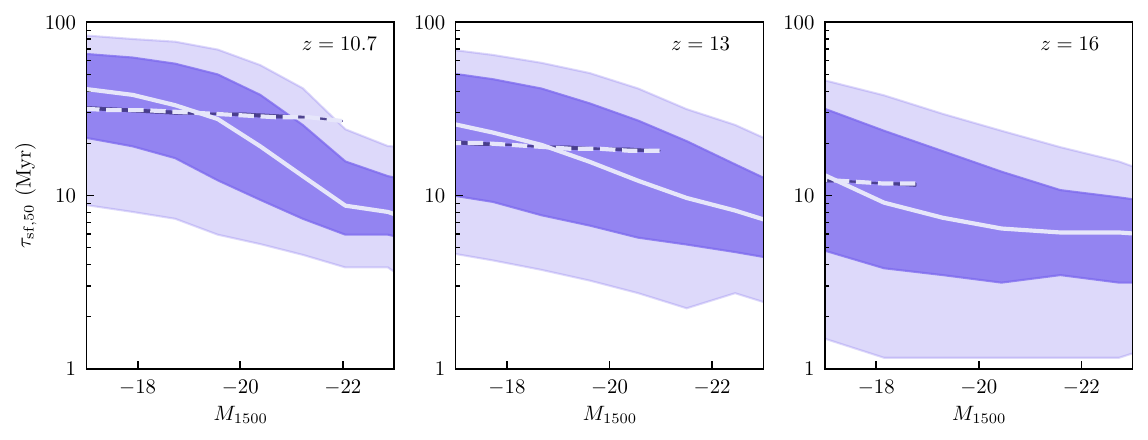}
  \caption[]{The median age of stars of model galaxies, $\tau_{\rm sf,50}$, in millions of years as a function of their $\lambda=1500\,{\rm \AA}$ absolute magnitude, $M_{1500}$ at redshifts $z=10.7, 13, 16$. The dashed lines show the median $\tau_{\rm sf,50}$ in the model without SFR stochasticity. The stellar ages in this model are quite small due to rapid growth of galaxies at these redshifts. The solid lines show the median $\tau_{\rm sf,50}$ in the model with SFR stochasticity that matches observed UV LF at $z>10$ (shown by solid lines in Fig. \ref{fig:lf_z1116}). The shaded regions show the $68.2$ and $95.4$ percentiles of the $\tau_{\rm sf,50}$ distribution.}
   \label{fig:m1500_t50}
\end{figure*}

An indicator of the star formation histories and stochasticity of the current star formation is the ratio of star formation rates averaged over the past 10 Myr and 100 Myr. Figure \ref{fig:sfr10_scatter} shows this ratio as a function of the absolute magnitude at $\lambda=1500\,\aa$ for model galaxies at $z=10.7$, $z=13$, $z=16$. The dashed lines in the figure panels show the median ratio in the model without SFR stochasticity which have values of $\rm SFR_{10}/SFR_{100}\approx 2$ at $z=10.7$ and $5$ at $z=16$. These values are due to the rapid growth and rapidly increasing SFR of model galaxies at these redshifts, which make most recent star formation larger than the star formation rate averaged over a longer period. 

Figure \ref{fig:sfr10_scatter} shows that SFR stochasticity results in a significant scatter around the median $\rm SFR_{10}/SFR_{100}$ ratio which is predicted to increase with increasing $z$. \cite{Cole.etal.2023} estimated the scatter of this ratio of $\approx 0.4-0.5$ dex at $9<z<12$, consistent with the scatter exhibited by galaxies in our stochastic SFR model. 
We note that the scatter is significant both for the values of $\rm SFR_{10}/SFR_{100}$ above and below the median value. SFR stochasticity thus produces galaxies with enhanced and depressed recent star formation at the same $M_{1500}$. This means there should be a wide range of ionizing luminosities produced by galaxies of a given UV luminosity. This prediction is potentially testable using observational estimates of the H$_\alpha$ and $L_{1500}$ luminosities. Figure \ref{fig:lha_luv_scatter} shows this ratio predicted by our model at the same three redshifts. The H$_\alpha$ luminosity was computed assuming Case B ionization as \citep[e.g., eq. 2 in][]{Kennicutt.1998}
\begin{equation}
    L_{\rm H_\alpha} = 1.37\times 10^{-12} \dot{N}_{\rm ion}\ \rm erg\, s^{-1}, 
\end{equation}
where $\dot{N}_{\rm ion}$ is the emission rate of photons with $\lambda<912\,\aa$ per second. This flux was computed self-consistently for model galaxies using their star formation histories and the tables of ionizing flux for single-age stellar populations with binaries from the BPASS stellar population synthesis package version 2.3 \citep{Byrne.etal.2022}.

In addition to modulating the $L_\alpha/L_{1500}$ ratio, the scatter in the ionizing flux should result in a wide range of line ratios for the ions sensitive to ionizing radiation, such as OII \citep{Kewley.etal.2019}. The highly bursty star formation will result in a wide range of the [OIII]/[OII] ratios in $z>10$ galaxies.

Finally, Figure \ref{fig:m1500_t50} shows that the median age of stars in model galaxies, $\tau_{\rm sf,50}$, is $\approx 10-40$ Myr at $z\in[11,16]$  broadly consistent with estimates for observed galaxies at these redshifts \citep[e.g.,][]{Franco.etal.2023,Morishita.Stiavelli.2023,Robertson.etal.2023,Robertson.etal.2024,Curtis_Lake.etal.2023,Carnall.etal.2023,Zavala.etal.2024}. Note that the median ages are $\approx 10-30$ Myr even without SFR stochasticity due to extremely rapid growth of galaxies at these redshifts. 

The SFR stochasticity introduces a distinct trend of decreasing median $\tau_{\rm sf,50}$ and decreasing scatter with increasing galaxy luminosity. This is because with the levels of SFR stochasticity required to match observed UV LF most of the bright galaxies become bright due to a strong burst of star formation, which reduces the median age of their stars. Indications of such a trend are found for galaxies at $z<9$ \citep{Endsley.etal.2023}.

These trends represent a potential test of the scenario in which UV LF function at $z>10$ is shaped by the significant SFR stochasticity.
Figure \ref{fig:m1500_t50} also shows that the scatter of stellar median ages is quite large at $M_{1500}\lesssim -20$ due to the variation of periods of specific bursts and lulls in star formation.

\subsection{Galaxy-halo connection}
\label{sec:galhalo}

Given that the model with SFR stochasticity reproduces the observed estimates of the UV LF across $z\in 11-16$, it is interesting to consider the galaxy-halo connection it implies in the form of the UV luminosity-halo mass relation. 

\begin{figure*}
  \centering
  \includegraphics[width=\textwidth]{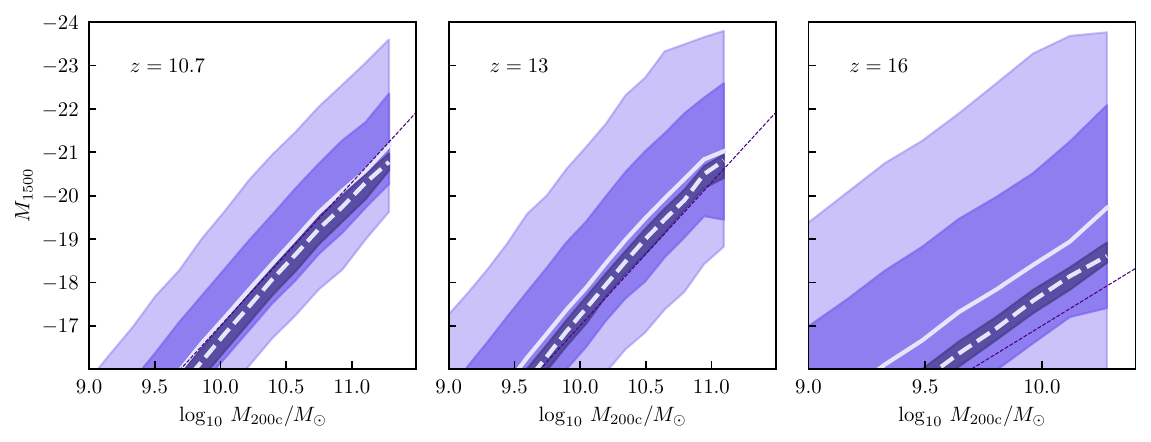}
  \caption[]{Absolute magnitude at $1500\,\aa$, $M_{1500}$, as a function of halo mass, $M_{\rm 200c}$, of model galaxies at $z=10.7, 13, 16$. The dashed lines show the results of the model with the fiducial parameters and {\it decreasing molecular depletion time}  with increasing $z$ from $\tau_{\rm sf}=1$ Gyr at $z=10.7$ to $0.25$ Gyr at $z=16$ and when stochasticity is not included. The scatter in this model shown by the narrow dark blue bands around the dashed lines is negligible (a small amount of scatter results from the different gas disk sizes assumed in the model for each galaxy). The solid lavender lines show the median relations for the same model but with SFR stochasticity added with the PSD parameters  $\alpha=2$, $\tau_{\rm break}=100$ Myr and $\sigma_\Delta=0.13, 0.15, 0.22$ at $z=10.7, 13, 16$. The UV LF for this model is shown by the solid lines in Figure~\ref{fig:lf_z1116}.}. 
   \label{fig:m200_m1500}
\end{figure*}

Figure \ref{fig:m200_m1500} shows the UV absolute magnitude at $\lambda=1500\,\aa$ as a function of halo mass $M_{\rm 200c}$ at redshifts $z=10.7, 14, 16$. For reference, the dotted line in each panel shows the relation 
\begin{equation}
M_{1500} = -17 - 3.3\left(\log_{10}M_{200}-10\right), 
\end{equation}
which approximates the median relation of the stochastic SFR model at $z=10.7$. 

\begin{figure*}
  \centering
  \includegraphics[width=\textwidth]{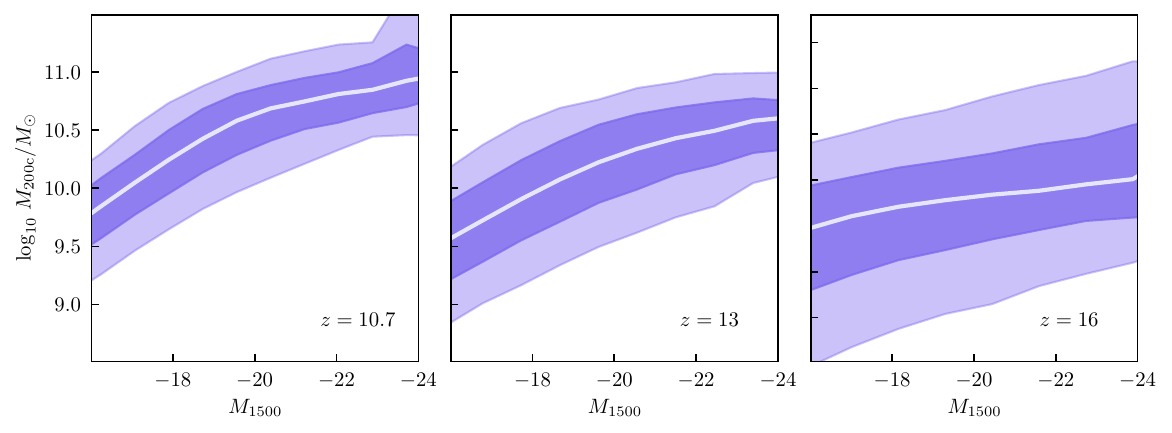}
  \caption[]{Distribution of halo mass, $M_{\rm 200c}$, of model galaxies as a function of absolute magnitude at $1500\, \aa$, $M_{1500}$, at $z=10.7, 13, 16$. The plots show the results of the model with the fiducial parameters and {\it decreasing molecular depletion time}  with increasing $z$ from $\tau_{\rm sf}=1$ Gyr at $z=10.7$ to $0.25$ Gyr at $z=16$ and SFR stochasticity added with the PSD parameters  $\alpha=2$, $\tau_{\rm break}=100$ Myr and $\sigma_\Delta=0.13, 0.15, 0.22$ at $z=10.7, 13, 16$ with which this model matches observed UV LF. The solid lavender lines show the median $M_{\rm 200c}$ in bins of $M_{1500}$, while the $68.2$ and $95.4$ percentiles of the distribution are shown by the darker and lighter shaded bands. The figure shows that most model galaxies in the observed luminosity range ($M_{1500}\lesssim -17$) are hosted by halos of a fairly narrow mass range. The average mass corresponding to observed luminosities decreases with increasing $z$. The increasing SFR stochasticity at higher $z$ leads to significant flattening of the relation.}. 
   \label{fig:m1500_m200}
\end{figure*}

The median luminosities at a given halo mass increase with increasing redshift and their scatter increases due to the assumed increase in the SFR stochasticity at higher $z$. The corresponding increase in the luminosity of galaxies in the model without SFR stochasticity is much milder. In the presence of significant scatter the distribution pdf $p(M_{1500}\vert M_{\rm 200c})$ is generally different from the pdf $p(M_{\rm 200c}\vert M_{1500})$ and Figure \ref{fig:m1500_m200} shows that this is the case here. The relation between the median $M_{\rm 200c}$ in bins of $M_{1500}$ and absolute magnitude is no longer linear and it becomes progressively flatter with increasing $z$. This flattening is the consequence of the lack of halos of mass $M_{\rm 200c}\gtrsim 3\times 10^{11}\, M_\odot$ that are rare at these redshifts.

Figure \ref{fig:m1500_m200} shows that most model galaxies in the observed luminosity range ($-21\lesssim M_{1500}\lesssim -17$) are hosted by halos in a narrow mass range of less than two orders of magnitude. The stellar masses of model galaxies in such halos are in the range  $M_\star\approx 10^6-5\times 10^9\, M_\odot$. The median stellar mass -- halo mass relation for this range of $M_\star$ is well described by a power law
\begin{equation}
    M_\star = f_\star\times 0.15\times 10^{10}\,M_\odot\left(\frac{M_{200}}{10^{10}\,M_\odot}\right)^{1.4}
\label{eq:msmh}
\end{equation}
where $f_\star$ is the fraction of the baryon budget that was converted into stars in halos of mass $M_{200}$.

\begin{figure*}
  \centering
  \includegraphics[width=0.99\textwidth]{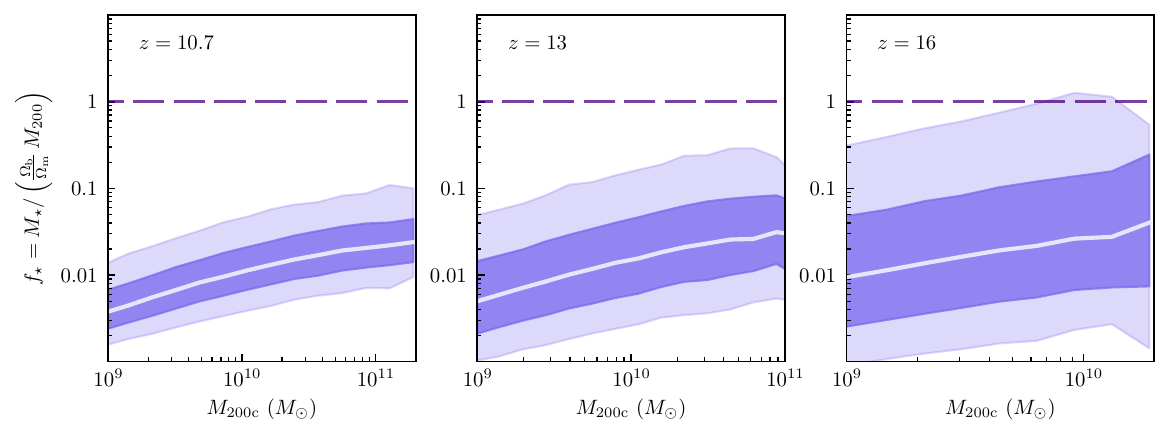}
  \caption[]{Fraction of the baryon mass that should be available to halos, $\Omega_{\rm b}/\Omega_{\rm m}\,M_{200}$,  converted into stars in model galaxies at redshifts $z=10.7$, $z=13$, and $z=16$. The solid lines show the median of the distribution, while dark and light-shaded regions show the $68.2\%$ and $95.6\%$ regions of the distribution of $f_\star$ at a given $M_{200}$. The horizontal dashed line shows $f_\star=1$ when galaxies convert their entire available baryon budget into stars.} 
   \label{fig:fstar_m200_z1116_dec_tauh2}
\end{figure*}

Figure \ref{fig:fstar_m200_z1116_dec_tauh2} shows that the median $f_\star$ varies from $f_\star\approx 0.005-0.02$ at $z=10.7, 12, 13$ to $f_\star\approx 0.01-0.05$ at $z=16$. Thus, halos are expected to convert only a few percent of their available baryons into stars on average to match observed UV LFs at $z>10$ in agreement with conclusions of \citet{Mason.etal.2023}. 

However, the scatter around the median increases dramatically with increasing redshift, as required by the condition to match the observational estimate using UV LFs. The tail of the distribution at $z=13$ reaches $f_\star\approx 0.1-0.3$. These values are still very reasonable and are comparable to the stellar fractions of $L_\star$ galaxies at $z=0$, such as the Milky Way. 
At $z=16$, however, the $95.4$ percentile of the distribution extends to $f_\star\approx 1$, which indicates that the SFR stochasticity of the model at this redshift is as high as it can be.

\section{Discussion}
\label{sec:discussion}

\subsection{Comparisons with previous studies and alternative scenarios to explain $z>10$ UV LF}

During the last two years, several scenarios have been put forth to explain the high abundance of UV-bright galaxies at $z>10$ \citep[see, e.g., Section 6.1 of][]{Harikane.etal.2023}.  

For example, \citet{Dekel.etal.2023} posited an idea that galaxies at these redshifts undergo starbursts in which usual feedback processes do not have the time to operate and gas is converted into stars in bursts with high efficiency.
Several studies have also proposed to explain the high abundance of UV bright galaxies by invoking an increase of star formation efficiency with increasing redshift \citep{Boylan-Kolchin.2023,Qin.etal.2023,Mason.etal.2023}. 

Our results indicate that increased star formation efficiency is not necessary to explain UV LF at $z\leq 13$ when SFR is stochastic, as the highest fractions of baryons converted into stars are reasonable $f_\star\lesssim 0.3$. Moreover, our model results show that if outflows are as efficient as they need to be to explain properties of $z=0$ dwarf galaxies higher star formation efficiency results in only a modest increase of UV LF amplitude. This is because higher star formation efficiency is partly compensated by higher gas outflow rate, which results in smaller gas mass. In simulations with weak feedback such regulation does not occur and higher UV luminosities can be achieved by increasing star formation efficiency. However, galaxies in such simulations may have incorrect properties at lower redshifts. 

Another discussed idea is the top-heavy IMF that increases the UV luminosity per unit stellar mass formed \citep{Inayoshi.etal.2022,Trinca.etal.2024,Yung.etal.2024,Ventura.etal.2024}. There are good reasons to expect top-heavy IMF at these high redshifts, where the minimum temperature of the ISM gas is higher due to higher cosmic microwave background temperature \citep[e.g.,][]{Jermyn.etal.2018,Chon.etal.2022} or other star-formation related processes \citep[e.g.,][]{Riaz.etal.2020,Steinhardt.etal.2022} and where a significant fraction of stars may be forming in massive bound clusters \citep[e.g.,][]{Belokurov.Kravtsov.2023,Adamo.etal.2024,Mowla.etal.2024}. The latter are environments conducive to the formation of massive and supermassive stars via stellar collisions \citep[e.g.,][]{Katz.etal.2015,Gieles.etal.2018,Katz.2019}. \citet{Rasmussen.Cueto.etal.2024}, however, explored evolving IMF within their galaxy formation model and found that top-heavy IMF by itself is unlikely to explain the bright end of UV LF at $z>10$ because the increase of the UV luminosity per unit mass of stars born is counteracted by the suppression of star formation due to increased feedback. 

The stochasticity of SFR was proposed as a natural explanation of the high amplitude of UV LF at $z>10$ in several recent studies \citep{Mason.etal.2023,Shen.etal.2023,Sun.etal.2023,Munoz.etal.2023}. 
Qualitatively, our results are consistent with their conclusions. However, we find that the amount of stochasticity required to match the observed UVLFs at $z=11-13$ is considerably smaller than was concluded in most previous studies.
\citet{Shen.etal.2023} presented results of a simple model assuming a constant efficiency of star formation and SFR proportional to the halo mass accretion rate. They concluded that to explain UV LF at $z\lesssim 9$ a modest level of scatter in $M_{\rm UV}$ of $\sigma_{M_{\rm UV}}\approx  0.75$ consistent with our finding. At $z>10$ \citet{Shen.etal.2023} deduce a much larger required scatter of $\sigma_{M_{\rm UV}}\approx 2.0$ at $z=12$ and $\sigma_{M_{\rm UV}}\approx 2.5$ at $z=16$. 
\citet{Mason.etal.2023}, on the other hand,  concluded that 
$\sigma_{M_{\rm UV}}\approx 1.5$ is sufficient to match UV LFs at $z\approx 11-13$, the number closer to our estimate. 

\begin{figure}
  \centering
  \includegraphics[width=0.49\textwidth]{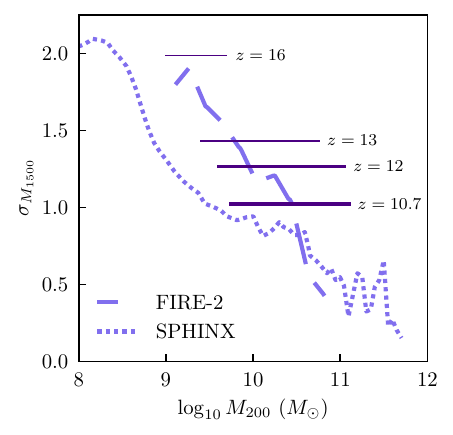}
  \caption[]{Standard deviation of the UV absolute magnitude at $\lambda=1500\,{\rm \AA}$, $\sigma_{M_{1500}}$ as a function of the logarithm of halo mass in the FIRE-2 simulations \citep{Sun.etal.2023} and in the SPHINX galaxy formation simulations \citep{Katz.etal.2023}. The solid horizontal lines show $\sigma_{M_{1500}}$ for our fiducial model that matches the observational estimates of UV LF at different redshifts. The horizontal extent of each line shows the $1\sigma$ spread of halo mass that host model galaxies contributing to UV LF probed by current observations at each $z$. }
   \label{fig:sigma_m1500}
\end{figure}

We show that at $z=11-13$ the observed UV LF can be reproduced with a considerably smaller scatter of $\sigma_{M_{\rm UV}}\approx 1-1.4$. As shown in Figure \ref{fig:sigma_m1500}, these values of $\sigma_{M_{\rm UV}}$ are consistent with the scatter measured in the FIRE-2 simulations. The figure shows the standard deviation of the UV absolute magnitude at $\lambda=1500\,\aa$, $\sigma_{M_{\rm UV}}$ as a function of the logarithm of halo mass in the FIRE-2 simulations \citep{Sun.etal.2023} and SPHINX galaxy formation simulations \citep{Katz.etal.2023}. The solid horizontal lines show $\sigma_{M_{\rm UV}}$ for our fiducial model that matches the observational estimates of UV LF at different redshifts with the extent of the lines showing the effective halo mass range contributing to galaxies in the observed luminosity range. 
Note that our model matching observational estimates of UV LF corresponds to $\sigma_{M_{\rm UV}}$ similar to those measured in the FIRE-2 simulations and we also reproduce the UV LF estimated using FIRE-2 simulations (see the $z=12$ panel of Figure \ref{fig:lf_z1116}). 

The scatter in the SPHINX simulation is consistent with the FIRE-2 simulations at $M_{\rm 200c}\gtrsim 3\times 10^{10}\,M_\odot$ but is smaller at smaller masses. At $z=10.7$ the level of $\sigma_{M_{\rm UV}}$ in the SPHINX is close to the level required to match observed UV LF, but is smaller at higher $z$. The scatter in the SPHINX does reach values of $\sigma_{M_{\rm UV}}\approx 1.5-2$ but at $\log_{10} M_{\rm 200c}\lesssim 9$. 
There are thus differences in the predicted degree of SFR stochasticity in current simulations, likely due to different implementations of star formation and feedback. 

At $z=16$ we find that $\sigma_{M_{\rm UV}}\approx 1.9-2$ is required by existing estimates of galaxy abundance at $-22\lesssim M_{\rm UV}\lesssim -20$. This is also smaller than the previous estimate of \citet{Shen.etal.2023}. We believe the difference arises because \citet{Shen.etal.2023} first compute the UV LF without scatter and then convolve it with a Gaussian of a given width. However, in the presence of a large scatter this procedure is not accurate and UV LF needs to be computed by a convolution of the halo mass function with the Gaussian pdf $p(M_{\rm UV}\vert M_{\rm 200c})$. If we convolve our $z=16$ UV LF in the model without SFR stochasticity with a Gaussian as was done by \citet{Shen.etal.2023} we do indeed find that $\sigma_{M_{\rm UV}}\approx 2.5$ is needed to match observational UV LF estimate. Yet, the average scatter in our fiducial model that matches observations at a given halo mass is only $\sigma_{M_{\rm UV}}\approx 1.9$.

In fact, we show in Section \ref{sec:galhalo} that $\sigma_{M_{\rm UV}}\approx 2$ at $z=16$ we deduce should be considered an upper limit because it requires that a small fraction of the model galaxies should form more stars than the mass of available baryons. This shows that higher levels of SFR stochasticity than those used in our model are unrealistic. 

\subsection{Other sources of scatter of UV luminosity}

SFR stochasticity is not the only process that can contribute to the scatter of UV luminosity.  
Fluctuations of $M_{\rm UV}$ can be enhanced for a given SFR stochasticity if the initial mass function of stars becomes top-heavy during SFR bursts. The surface density of star formation, $\Sigma_{\rm SFR}$ should increase dramatically during such bursts and it is known that in low-$z$ galaxies the fraction of stars formed in massive bound clusters increases with increasing $\Sigma_{\rm SFR}$ \citep[e.g., see][for a review]{Adamo.etal.2020rev}. At $\Sigma_{\rm SFR}\gtrsim 10-30\, M_\odot\,\rm yr^{-1}\,kpc^{-2}$, characteristic of extreme starbursts at $z=0$ and observed in a significant fraction of galaxies at $z>10$ \citep[e.g.,][]{Morishita.Stiavelli.2023,Robertson.etal.2023,Morishita.etal.2024}, $\gtrsim 50\%$ of star formation is expected to form in massive bound clusters according to the observed trend for local galaxies. As $\Sigma_{\rm SFR}$ can increase to even higher values during strong SFR bursts (in fact $\Sigma_{\rm SFR}$ for many $z>10$ are lower limits due to limited angular resolution of observations), close to $100\%$ of stars may form in massive bound clusters.   

This is now directly observed in galaxies at $z\gtrsim 7$ \citep{Adamo.etal.2024,Mowla.etal.2024,Castellano.etal.2024} and in the low-metallicity stellar population of the Milky Way that should have formed at similar redshifts \citep{Belokurov.Kravtsov.2023}. As noted above, massive bound clusters are environments conducive to the formation of massive and supermassive stars \citep[e.g.,][]{Gieles.etal.2018}. When close to $100\%$ of stars form in massive clusters the IMF can become top-heavy due to the formation of such stars. 
The impact of the SFR bursts on the UV luminosity can thus be considerably enhanced compared to the standard IMF assumed in our analysis. 
On the observational side, potential signatures of top-heavy IMF have been indeed indicated by recent observational analyses \citep{Cameron.etal.2024}. 

\subsection{Observational indications of bursty star formation at $z>5$}
\label{sec:tests}

A growing number of observational analyses indicate a large degree of burstiness in $z>7$ galaxies. \citet{Endsley.etal.2023} show that the population of galaxies at $z\approx 7-9$ exhibits large fluctuations of SFR and that brightest galaxies tend to be undergoing a starburst. 
\citet{Simmonds.etal.2024} show that bursty dwarf galaxies at $4\lesssim z\lesssim 9$ contribute more ionizing photons than non-bursty galaxies of the same stellar mass and thus likely play a key role in reionizing the Universe. Moreover, they find strong evidence that ionizing luminosity increases with decreasing galaxy luminosity and increasing redshift indicating that burstiness increases with increasing $z$.

Likewise, \citet{Ciesla.etal.2023} used JWST JADES survey data to estimate the amount of SFR stochasticity in galaxies at $6<z<12$ and found that stochasticity is not only ubiquitous but is increasing towards higher redshifts \citep[see also][]{Cole.etal.2023}. They find that at $z>9$ up to $\approx 90\%$ of galaxies of mass $M_\star\gtrsim 10^9\, M_\odot$ show signs of bursty star formation rate in the past 100 Myr, while this fraction drops to $\approx 15\%$ at $z<7$. This is consistent with our finding that the amount of SFR stochasticity required to match the UVLFs at $z>5$ decreases rapidly around $z\approx 10$. 

\citet{Ciesla.etal.2023} also estimated that the dispersion of UV luminosity $\sigma_{M_{\rm UV}}$ stays approximately constant as a function of $z$ at $\sigma_{M_{\rm UV}}\approx 1.2$. Thus, their results imply that it is the fraction of starbursting galaxies that rapidly declines with decreasing redshift, not the dispersion $\sigma_{M_{\rm UV}}$. We note, however, that at $z>9$ where almost all galaxies are bursty according to their analysis, our results indicate that $\sigma_{M_{\rm UV}}\approx 1.2$ is sufficient to match the observed UVLF at $z\lesssim 13$, especially if lower molecular gas depletion time is considered (see Figure~\ref{fig:lf_z1116}).



\subsection{Observational signatures of bursty star formation}

Our results show that SFR stochasticity at the level required to match UV LF estimates at $z>10$ has several potentially testable predictions. First, this scenario predicts that UV LF does not steepen with increasing $z$ at $z>10$, but gets even somewhat shallower at $z>12$ (see Section \ref{sec:uvlf} and Figures \ref{fig:lf_z1116} and \ref{fig:lf_flattening}). This prediction should be testable with future JWST observations. 

The model also predicts a decrease in the median stellar age of the stellar population with increasing UV luminosity \citep[see also][]{Ren.etal.2019,Mason.etal.2023}. A significant scatter of median stellar ages for galaxies of $M_{1500}\gtrsim -21$ is expected with the significant decrease of scatter for brighter galaxies (see Fig. \ref{fig:m1500_t50}). The increasing SFR stochasticity at $z>10$ should result in a larger fraction of galaxies with extremely young ages $\lesssim 5-10$ Myr. 

Significant SFR stochasticity modifies the distribution of virial masses of halos that host observed galaxies within a given $M_{1500}$ range (see Fig. \ref{fig:m1500_m200}). In this scenario, UV-bright galaxies occupy predominantly low-mass halos and become UV bright during SF bursts. In contrast, in the scenario in which the UV bright galaxies are bright due to high SF efficiency, the bright galaxies should occupy massive halos. As pointed out by \citet{Munoz.etal.2023}, this difference should in principle be manifested in the different clustering amplitude of UV-bright galaxies predicted by these two scenarios. 

SFR stochasticity results in a substantial scatter of the star formation rates averaged over 10 and 100 Myr, SFR$_{10}$/SFR$_{100}$ (Fig. \ref{fig:sfr10_scatter}). The scatter of this ratio in model galaxies is dominated by the fluctuations of SFR$_{10}$, which is in qualitative agreement with measurements of \citet{Cole.etal.2023} at $z\approx 10-11$. To explain the observational estimates of UV LF at higher $z$, SFR stochasticity must increase at $z>11$ and this leads to the increase 
the scatter around the median SFR$_{10}$/SFR$_{100}$ from $\approx 0.5$ dex at $z=10.7$ to $\approx 0.7$ dex at $z=13$ and $\approx 1$ dex at $z=16$. 

A related quantity is the ratio of H$_\alpha$ to UV luminosity (Fig. \ref{fig:lha_luv_scatter}). This ratio is essentially the ionization production efficiency, $\xi_{\rm ion}$:
\begin{equation}
    \xi_{\rm ion} = \frac{\dot{N}_{\rm ion}}{L_{1500}}\ {\rm erg^{-1}\, Hz}\approx \frac{7.28\times 10^{11}\,L({\rm H}_\alpha)}{L_{1500}}\ \rm erg^{-1}\, Hz
\end{equation}
where $\dot{N}_{\rm ion}$ is the number of the Lyman continuum photons emitted by a galaxy per second, and $L_{1500}$ is the monochromatic luminosity at $\lambda=1500\,{\rm \AA}$ in units of $\rm erg\,s^{-1} Hz^{-1}$. Note that the $L({\rm H}_\alpha)/L_{1500}$ ratio predicted by our model assumes the standard IMF (the IMF adopted in BPASS v2.3). This ratio should be different if top-heavy IMF is used to explain the high UV luminosities of $z>10$ galaxies. 

The fluctuations of $\xi_{\rm ion}$ due to SFR burstiness will result in the fluctuations of the ionization factor $U$, which is proportional to $\dot{N}_{\rm ion}$.  The stochastic SFR scenario predicts that the brightest galaxies $M_{1500}\lesssim -21$ should have larger H$_\alpha$ equivalent widths due to the smaller median age of their stellar population. The ratio of H$_\alpha$ to UV luminosities should have a smaller scatter than in fainter galaxies (see Fig. \ref{fig:lha_luv_scatter}). Indications of such trend have been found in $z<9$ galaxies \citep{Endsley.etal.2023}.

The fluctuations of the ionization factor will also result in fluctuations of the line flux rations for the lines sensitive to ionizing radiation, such as [OIII]/[OII], [NIII]/[NII], etc \citep[see, e.g., Fig. 7 in][]{Kewley.etal.2019}. Given that SFR stochasticity is required to increase with redshift the effects of the ionizing flux fluctuations should also increase. We thus can expect an increase in the typical [OIII]/[OII] line flux ratios with increasing $z$. 

Also, strong fluctuations of SFR and corresponding ionizing flux are expected to lead to suppression of $158\,\mu\rm m$ [CII] emission in high-$z$ galaxies \citep{Katz.etal.2023b}, which means that increasing SFR burstiness should result in a lower number of [CII] detections at $z>10$ in ngVLA observations \citep{Carilli.etal.2018}. \citet{Katz.etal.2023b} show that combined information from the [CII], H$_\alpha$, and $L_{1500}$ can be used to identify galaxies that experienced extreme SFR fluctuations recently.

\section{Summary and conclusions}
\label{sec:conclusions}

We use the \texttt{GRUMPY} galaxy formation framework to predict stellar masses, star formation rates, and UV luminosities of galaxy populations at $z\in 5-16$.  
The galaxy formation model was previously shown to reproduce observed properties of $\lesssim L_\star$ galaxies at $z=0$ down to the faintest ultra-faint dwarf galaxies \citep[][]{Kravtsov.Manwadkar.2022,Manwadkar.etal.2022,Kravtsov.Wu.2023}. Here we use the model with halo mass tracks obtained by integration of a new simple approximation to the halo mass accretion rate accurate across a wide redshift interval of $0<z<20$ (eq. \ref{eq:dmdt}). We add SFR stochasticity to the star formation rate predicted by the model to explore its effects on the observed UV LF in a controlled manner and its potential observational implications. Our results and conclusions are as follows. 

\begin{itemize}
    \item[1.] We show that the model can match observational estimates of observed UV LFs at $5<z<10$ with a modest level of SFR stochasticity resulting in $\sigma_{M_{\rm UV}}\approx 0.75$ (Figure \ref{fig:lf_z59}). Such SFR stochasticity is consistent with the expectation of high-resolution galaxy formation simulations.\\
    
    \item[2.] To match the observed UV LFs at $z\approx 11-13$ the SFR stochasticity should increase by a factor of $\approx 1.5-2$ producing $\sigma_{M_{\rm UV}}\approx 1-1.3$ (see Table \ref{tab:sfr_stoch_params}). We show that such a stochasticity level is consistent with the SFR stochasticity observed in high-resolution zoom-in simulations and is significantly smaller than the level of stochasticity deduced by previous studies. \\

    \item[3.] We show that if the SFR stochasticity increases with increasing redshift at the rate suggested by our results, the observed UV LF at $z>10$ does not steepen and even becomes somewhat flatter at $z>12$ (Figures \ref{fig:lf_z1116} and \ref{fig:lf_flattening}). \\
        
    \item[4.] We show that our model matching $z>10$ UV LFs is in good agreement with existing measurements of the SFR and stellar masses of galaxies at these redshifts (Figure \ref{fig:ms_sfr_z1116_dec_tauh2}). The model also predicts the scatter of the star formation rates averaged over 10 and 100 Myr, SFR$_{10}$/SFR$_{100}$ in agreement with the current estimate at $z\approx 10-11$. This scatter is predicted to increase at higher $z$ due to the increase of SFR stochasticity required to match the observational estimates of UV LFs at $z>10$ (Figure \ref{fig:sfr10_scatter}).\\ 
    
    \item[5.] The typical median age of stellar populations in the model with SFR stochasticity is $\approx 20-30$ Myr with a large scatter and the model predicts a systematic decrease of the median stellar age and scatter around it at the absolute magnitudes of $M_{1500}\lesssim -21$ (see Figure \ref{fig:m1500_t50}).\\ 

    \item[6.] The scatter of short-term SFR and stellar ages results in a substantial scatter in the ionizing flux at a given UV luminosity. This has several consequences that can be potentially probed by observations. For example, the increase of SFR stochasticity with redshift results in a larger fraction of galaxies experiencing a strong burst or a significant lull in their star formation. The associated fluctuations of the ionization parameter will result in the increasing scatter in the line fluxes and their ratios with increasing $z$ for the lines sensitive to this parameter (see Discussion in Section \ref{sec:tests}). \\
    
    \item[7.] According to our model, galaxies of $M_{\rm UV}\in [-17, -22]$ are hosted by halos of mass $M_{\rm 200c}\approx 5\times 10^{9}-3\times 10^{11}\,M_\odot$ at $z\approx 10.7$ but are hosted in halos of progressively smaller mass at higher $z$. \\
    
    \item[8.] Stellar masses of model galaxies at observed luminosities at $z>10$ are in the range $\approx 5\times 10^6- 5\times 10^9\, M_\odot$ with the median stellar mass-halo mass relation well described by 
    $M_\star = f_\star\times 0.15\times 10^{10}\,M_\odot M_{200,10}^{1.4}$, where $M_{200,10}$ is halo mass in units of $10^{10}M_\odot$ and $f_\star$ is the fraction of the baryon budget that was converted into stars in halos of mass $M_{200}$, which varies from $f_\star\approx 0.005-0.03$ at $z=10.7-13$ to $f_\star\approx 0.01-0.05$ at $z\approx 16$ (Figure \ref{fig:fstar_m200_z1116_dec_tauh2}). On average the model galaxies thus convert only a few percent of the available baryon budget into stars and extremely high star formation efficiency is not required to match observed UV LF at $z\leq 13$.\\

    \item[9.] The scatter of $f_\star$ is substantial but most galaxies have $f_\star\lesssim 0.4$ at $z\approx 11-13$. At $z=16$, on the other hand, the level of SFR stochasticity required to match the current estimate of the abundance of UV bright galaxies predicts that a small fraction of galaxies have $f_\star>1$ (Figure \ref{fig:fstar_m200_z1116_dec_tauh2}). Thus, the level of SFR stochasticity required to match current estimates of UV LF at $z\approx 16$ are as high as they can be.\\

\end{itemize}

Our results and results of previous related theoretical studies strongly motivate exploration of the physical processes that can drive the burstiness of star formation in both dwarf and massive galaxies at $z\gtrsim 5-10$ as these processes may be prevalent during early stages of galaxy evolution and can thus shape the properties of low-metallicity stellar populations of galaxies. On the observational side, several different observational probes discussed in Section \ref{sec:tests} can be used to measure or constrain the level of SFR stochasticity at $z>10$, thereby providing a test of this explanation for high abundance of UV-bright galaxies at these redshifts.

\section*{Acknowledgments}

We are grateful to Brant Robertson, Harley Katz, and the UChicago structure formation group for useful discussions during this project. We are also grateful to Massimo Ricotti for sharing star formation histories of the simulated galaxies from \citet{Garcia.etal.2023} with us. 
AK was supported by the National Science Foundation grants AST-1714658 and AST-1911111 and NASA ATP grant 80NSSC20K0512.

We use simulations from the FIRE-2 public data release \citep{Wetzel.etal.2023}. The FIRE-2 cosmological zoom-in simulations of galaxy formation are part of the Feedback In Realistic Environments (FIRE) project, generated using the Gizmo code \citep{hopkins15} and the FIRE-2 physics model \citep{Hopkins.etal.2018}.
Analyses presented in this paper were greatly aided by the following free software packages: {\tt NumPy} \citep{NumPy}, {\tt SciPy} \citep{scipy}, {\tt Matplotlib} \citep{matplotlib}, {\tt FSPS} \citep{fsps} and its Python bindings package {\tt Python-FSPS}\footnote{\href{https://github.com/dfm/python-fsps}{\tt https://github.com/dfm/python-fsps}}, {\tt BPASS} stellar population synthesis tables for ionizing luminosity \cite{Byrne.etal.2022}, and {\tt Colossus} cosmology package \citep{colossus}. We have also used the Astrophysics Data Service (\href{http://adsabs.harvard.edu/abstract_service.html}{\tt ADS}) and \href{https://arxiv.org}{\tt arXiv} preprint repository extensively during this project and the writing of the paper.

\section*{Data Availability}
The codes and data used in this study are available upon request. 

\bibliography{references}

\appendix
\section{High-$z$ UV LF and variations of depletion time}
\label{app:lf_taudep}
\begin{figure}
  \centering
  \includegraphics[width=0.5\textwidth]{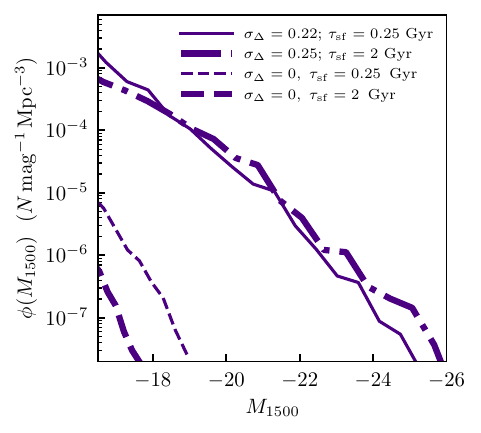}
  \caption[]{Model UV luminosity functions at redshift $z=16$. The dashed lines show model of UV LF with the fiducial parameters without any SFR stochasticity in the $\tau_{\rm sf}=2$ Gyr (thick dashed line) and  $\tau_{\rm sf}=0.25$ Gyr (thin dashed line). The thick dot-dashed and thin solid lines show results for the same models but with SFR stochasticity added with the PSD parameters  $\alpha=2$, $\tau_{\rm break}=100$ Myr and $\sigma_\Delta=0.25$ and 0.22, respectively (corresponding to $\sigma_{\rm M_{\rm UV}}=2.2$ and $1.9$, see Table \ref{tab:sfr_stoch_params}). The figure shows that the amplitude of UV LF can be increased with increasing efficiency of star formation (smaller $\tau_{\rm sf}$) or with SFR stochasticity. For more efficient star formation smaller amplitude of SFR stochasticity is required to match UV LF. }
   \label{fig:lf_z16}
\end{figure}

%
Star formation depletion time is one of the parameters of galaxy formation model used in this study. 
Figure \ref{fig:lf_z16} illustrates the effect of changing depletion time on the model UV LF. It compares the $z=16$ UV LF in the models with $\tau_{\rm sf}=2$ Gyr and $\tau_{\rm sf}=0.25$ Gyr without SFR stochasticity (thick and thin dashed lines, respectively). Although the amplitude of the LF is increased by an order of magnitude in the model with $\tau_{\rm sf}=0.25$ Gyr, its amplitude is still more than two orders of magnitude smaller than the corresponding models with SFR stochasticity at the level required to match the observational estimate of UV LF at this redshift (solid and dot-dashed lines). Thus, in our model LF at $z\gtrsim 13$ cannot be matched by a realistic increase of star formation efficiency. 
At the same time, the fiducial model that matches observational estimates of UV LF is consistent with the inferred star formation rates of $z>10$ galaxies. 

%

\section{Tests of the halo mass accretion rate approximation}
\label{app:mhacc}

In our model calculations, we use halo tracks constructed by integrating a simple expression for the halo mass accretion rate of equation \ref{eq:dmdt}. Here we present a comparison of this expression with other calibrations of halo mass accretion rate and demonstrate its accuracy. 

\begin{figure}
  \centering
  \includegraphics[width=0.49\textwidth]{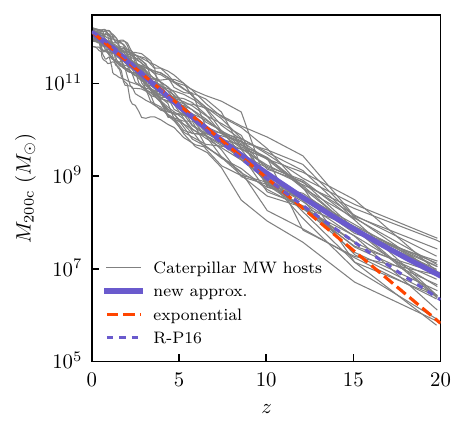}
  \caption[]{Comparison of the halo mass evolution obtained by integration of the approximation used in this study (eq.~\ref{eq:dmdt}) and individual mass accretion histories of the MW-sized halos from the Caterpillar simulation suite \citet{Griffen.etal.2016}. The dashed line shows exponential mass accretion rate approximation of \citet[][dashed line]{Correa.etal.2015}. }
   \label{fig:mahtest_cat}
\end{figure}

\begin{figure*}
  \centering
  \includegraphics[width=\textwidth]{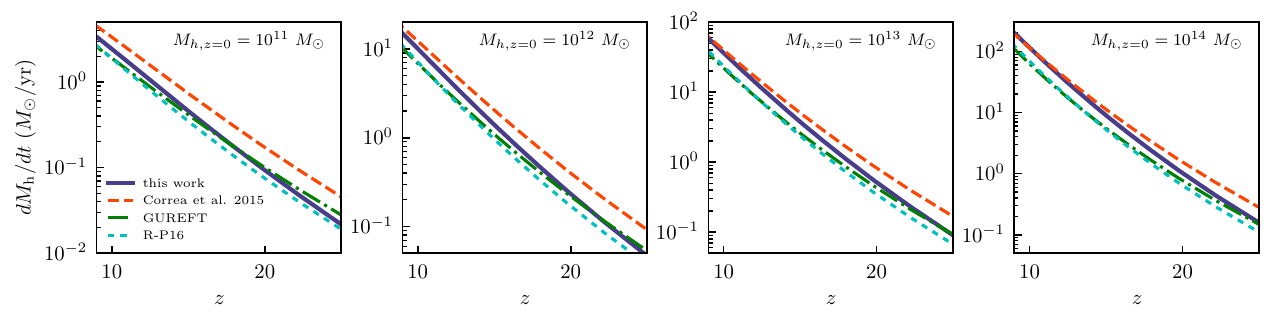}
  \caption[]{Comparison of the halo mass accretion rates using approximation used in this study (eq.~\ref{eq:dmdt}) and approximations of \citet[][dashed line]{Correa.etal.2015}, \citet[][dotted line]{Rodriguez.Puebla.etal.2016} and \citet[][dot-dashed line]{Yung.etal.2024} for the mean halo evolution tracks computed using integration of eq.~\ref{eq:dmdt} for halos with $z=0$ $M_{\rm 200c}$ mass of $10^{11}$,  $10^{12}$,  $10^{13}$,  $10^{14}$ in panels from left to right. }
   \label{fig:dmdt_comp}
\end{figure*}

Figure \ref{fig:mahtest_cat} shows halo tracks obtained by integrating different approximations of the halo mass accretion rate for halos of $M_{\rm 200c}=10^{12}\, M_\odot$ at $z=0$. Specifically, we use the mass accretion rate of equation \ref{eq:dmdt} and approximations of \citet{Rodriguez.Puebla.etal.2016} and the mass accretion rate of \citet{Correa.etal.2015} corresponding to the exponential approximation to the halo mass as a function of $z$ \citep{Wechsler.etal.2002}. The $M_{\rm 200c}(z)$ obtained with these approximations for $\dot{M}_{\rm 200c}$ is compared to the halo mass tracks for halos in the Caterpillar suite of high-resolution simulations of $M_{\rm 200c}\approx 10^{12}\, M_\odot$ halos \citep{Griffen.etal.2016}. All approximations produce similar tracks at $z>10$, but the approximation of the equation \ref{eq:dmdt} provides a better description of the mass evolution at higher $z$.

Figure \ref{fig:dmdt_comp} shows a direct comparison of the $\dot{M}_{\rm 200c}$ for halos of a broad range of $z=0$ mass to the halo mass accretion rate approximation of \citet{Correa.etal.2015}, \citet{Rodriguez.Puebla.etal.2016}, and the recent calibration of \citet{Yung.etal.2024b} focusing specifically on halo evolution at $z>5$. The figure shows that our approximation is close to the approximations of \citet{Rodriguez.Puebla.etal.2016} and \citet{Yung.etal.2024}.

\begin{figure}
  \centering
  \includegraphics[width=0.49\textwidth]{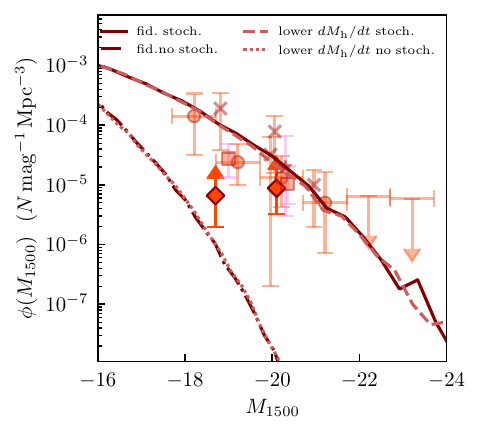}
  \caption[]{Comparison of the fiducial model prediction for the UV LF at $z=12$ with and without SFR stochasticity (long-dashed and solid line) with the model calculations in which halo mass accretion rate was reduced by 30\% at all redshifts (short-dashed and dotted lines), which corresponds approximately to the typical difference of our approximation for $\dot{M}_{\rm 200c}$ (eq. \ref{eq:dmdt}) and approximations of \cite{Yung.etal.2024} and \citet{Rodriguez.Puebla.etal.2016} at these $z$. The figure shows that differences of this level have a negligible effect on the UV LF at $z\sim 12$.}
   \label{fig:uvlf_z12_dmdt_comp}
\end{figure}

The differences of $\lesssim 30-40\%$ at $z<17$ exist but they have a small effect on the UV LF prediction. Figure \ref{fig:uvlf_z12_dmdt_comp} shows a comparison of the UV LFs at $z=12$ for our fiducial model and model in which the mass accretion rate was lowered by 30\% at all redshifts. The figure shows that the differences induced by such modifications of the halo mass accretion rate have a negligible effect on the UV LF prediction. Although this may seem surprising given the common assumption that $L_{1500}\propto {\rm SFR}\propto \dot{M}_{\rm 200c}$ in the calculations of UV LF for a halo population, in our model the star formation rate is set by the amount of star-forming gas, which constitutes only a fraction of the total gas and is not directly linked to the instantaneous halo mass accretion rate. This indicates that residual uncertainties in the halo mass accretion rate are not a significant source of uncertainty in galaxy formation model predictions at these redshifts. 

\label{lastpage}

\end{document}